\documentclass[11pt,a4paper]{article}

\usepackage[margin=1in]{geometry}
\usepackage[T1]{fontenc}
\usepackage[utf8]{inputenc}
\usepackage{mathptmx}
\usepackage{amsmath,amssymb}
\usepackage{graphicx}
\usepackage{booktabs}
\usepackage{placeins}
\usepackage[colorlinks=true,linkcolor=blue,citecolor=blue,urlcolor=blue]{hyperref}
\hypersetup{
  pdftitle={Optimal Stratification of a Sampling Frame: A Comparative Study of Classical, Quantum, and Quantum-Inspired Approaches},
  pdfauthor={Marco Ballin, Giulio Barcaroli},
  pdfkeywords={optimal stratification; sample allocation; Bethel-Chromy algorithm; SamplingStrata; quadratic unconstrained binary optimisation (QUBO); quantum annealing; Quantum Approximate Optimisation Algorithm (QAOA); quantum-inspired computing}
}
\usepackage{url}
\usepackage[round,authoryear]{natbib}
\usepackage{titlesec}
\usepackage{enumitem}
\setlist{topsep=4pt,itemsep=2pt}

\titleformat{\section}{\normalfont\Large\bfseries}{\thesection}{1em}{}
\titleformat{\subsection}{\normalfont\large\bfseries}{\thesubsection}{1em}{}

\usepackage{authblk}

\title{\textbf{Optimal Stratification of a Sampling Frame:}\\[4pt]
	\textbf{A Comparative Study of Classical, Quantum, and Quantum-Inspired Approaches}}
\author[1]{Marco Ballin}
\author[2]{Giulio Barcaroli}
\affil[1]{Istituto Nazionale di Statistica, \texttt{marco.ballin@istat.it}}
\affil[2]{Independent Consultant, \texttt{gbarcaroli@gmail.com}}
\date{}

\begin{document}
\maketitle

\textbf{Reproducibility Statement.} The data and software required to reproduce the computational workflow are available at \url{https://github.com/barcaroli/QuantumComputing}. 

\textbf{Conflict of Interest Statement.} The authors declare that they have no conflicts of interest.

\textbf{Funding Statement.} This research received no specific grant from any funding agency.

\textbf{Statement of Artificial Intelligence Use.} Artificial intelligence (AI) tools, including ChatGPT (OpenAI) and Claude (Anthropic), were used to generate the code and improve language, readability, and clarity of the paper. All AI-generated outputs were reviewed, verified, and edited by the authors. The authors are solely responsible for the scientific content presented in this work.

\bigskip

\begin{abstract}
\noindent Optimal stratification aggregates a large number of atomic strata into a small number of final strata so as to minimise the total sample size required to meet target precision constraints, an objective that is combinatorial and, once reformulated as a within-cluster dispersion surrogate, expressible as a quadratic unconstrained binary optimisation (QUBO) problem. This paper reports a comparative case study of four solvers for that surrogate, run under heterogeneous free-tier constraints, on an identical twenty-stratum frame drawn from the swissmunicipalities dataset: a D-Wave-formulated QUBO, solved in this study by classical simulated annealing because QPU access was unavailable, a gate-based quantum processor (IBM Quantum, running the Quantum Approximate Optimisation Algorithm), a photonic entropy-quantum-computing device (QCi Dirac-3), and a classical GPU-based Ising machine used as a control (Fixstars Amplify AE). The solvers differ not only in hardware but also in computational budget, iteration count and, for IBM, the encoding itself, so the comparison isolates a case study rather than a fully controlled experiment. Compared with a classical genetic-algorithm benchmark of 129 sample units, the best-known QUBO solution found in this study has an objective value of 501.0 and maps, after classical Bethel-Chromy evaluation, to a sample size of 160: an observed downstream gap of roughly twenty-four percent, associated with the best solution found rather than a certified optimum on either side. IBM's gate-based processor and the vendor-described photonic  entropy-quantum-optimisation device (Dirac-3) fall further short of even that value, at 262 (IBM) and 286 (Dirac-3); the gate-based result is consistent with degradation from SWAP-routing depth on a sparse qubit lattice, though this was not isolated from other possible contributors, while the photonic device, although it does perform a genuine optimisation, settles well short of the best-known value that a free graphics processor reaches in seconds. The paper concludes that, at the scale tested, a classical genetic algorithm remains the method of choice for optimal stratification. This conclusion reflects two main factors: (a) the limitations of the surrogate objective used in the QUBO-based approaches (a population-weighted within-cluster pairwise dispersion criterion) instead of the Bethel–Chromy sample-size objective, which cannot be directly expressed as a low-degree polynomial and therefore cannot be straightforwardly encoded as a QUBO; and (b) the solver and platform-specific limitations observed in the IBM and Dirac-3 runs. These findings are empirical and do not establish global optimality for either the genetic-algorithm result or the QUBO solutions.
\end{abstract}

\section{Introduction}
\label{sec:intro}
Optimal stratification, the problem of partitioning a sampling frame into strata so as to minimise the sample size needed for a target level of precision, is a long-standing and still actively used tool in survey methodology \citep{cochran1977}. Modern software solves it well: genetic algorithms, exemplified by the R package SamplingStrata \citep{barcaroli2014,ballin2013}, routinely find near-optimal aggregations of large frames of atomic strata, and are the de facto standard in official statistics.

Independently, the last decade has seen quantum annealers, gate-based quantum processors and photonic quantum-inspired devices become accessible, at limited scale, through commercial cloud services. Because many combinatorial optimisation problems, clustering among them, can be recast as quadratic unconstrained binary optimisation (QUBO) or, equivalently, Ising Hamiltonians \citep{lucas2014}, it is natural to ask whether these devices offer any advantage for optimal stratification specifically, and, if not, what precisely stands in the way.

Answering that question honestly requires more than running a solver and reporting a number. It requires (i) an explicit account of what each device is actually asked to minimise, since the true surveydesign objective, the Bethel sample size \citep{bethel1989,chromy1987}, is not itself expressible as a low-degree polynomial and every quantum or quantum-inspired formulation therefore substitutes a surrogate; (ii) a control that isolates the cost of that surrogate from the cost of the hardware; and (iii) a common frame, target and precision constraint across platforms, so that differences in outcome are not attributable to differences in the problem instance itself, even where computational budgets and encodings differ. Existing demonstrations of quantum optimisation rarely provide all three at once.

This paper provides a comparative case study along these lines, under the heterogeneous constraints that free-tier access to each platform actually imposes. Four solvers, a D-Wave-formulated QUBO solved in this study by classical simulated annealing because QPU access was unavailable, a gate-based quantum processor (IBM Quantum, running the Quantum Approximate Optimisation Algorithm), a photonic entropy-quantum-computing device (QCi Dirac-3), and a classical GPU-based Ising machine used as a control (Fixstars Amplify AE), are run on an identical twenty-stratum frame drawn from the swissmunicipalities dataset, against the same classical genetic-algorithm benchmark. Computational budgets, iteration counts and, for IBM, the encoding itself are not uniform across solvers, so observed differences cannot be attributed to hardware alone; a table making these differences explicit is given in Section~\ref{sec:case-study}. The classical control is still what makes the comparison informative: because it minimises the same one-hot QUBO used for D-Wave and Dirac-3 without the noise, embedding or circuit-depth limitations of physical quantum hardware, it helps distinguish how much of the observed gap is associated with that surrogate formulation as opposed to the solver executing it, without identifying an exact causal quantity. The IBM implementation uses a different binary-label encoding of the same underlying dispersion criterion and is therefore compared at the level of the resulting surrogate objective rather than term by term.

The remainder of the paper is organised as follows. Section~\ref{sec:problem} sets out the optimal-stratification problem formally. Section~\ref{sec:classical} describes the classical benchmark, the genetic algorithm implemented in SamplingStrata. Section~\ref{sec:quantum} introduces the QUBO and Ising formalism and describes, for each of the four solvers, how the stratification problem is mapped onto it and how much of the resulting computation is genuinely quantum. Section~\ref{sec:case-study} describes the case study and explains why the common experimental frame is limited to twenty atomic strata, a limit imposed by the free-tier constraints of the platforms tested rather than by the statistical problem itself. Section~\ref{sec:comparative} reports the comparative results and decomposes the gap between each solver and the classical benchmark into a formulation component and a solver/platform component. Section~\ref{sec:beyond-free-tier} considers what paid access to each platform would, and would not, change. Section~\ref{sec:wider-landscape} draws out the implications for hardware not yet tested, in particular trapped ions, and identifies the single further experiment judged most informative; a fuller survey of other platforms considered is given in Appendix~\ref{app:platforms}. Section~\ref{sec:conclusions} concludes.

\section{Problem definition: optimal stratification of a sampling frame}
\label{sec:problem}
Stratified sampling is a cornerstone of survey methodology \citep{cochran1977}. Given a finite population (a sampling frame) of N units, the units are partitioned into H mutually exclusive strata, and an independent sample is drawn from each stratum. When the strata are internally homogeneous with respect to the survey target variables, stratification lowers the variance of the estimates, so that a given precision can be reached with a smaller total sample size, or a smaller sample achieves a given precision. The design problem is therefore to choose the stratification that minimises the total sample size subject to precision constraints on the estimates of interest.

In the setting considered here the frame carries a small number of categorical auxiliary variables, obtained by discretising continuous covariates. The Cartesian product of the categories of these auxiliary variables defines a set of atomic strata: the finest stratification that the available covariates allow. Each atomic stratum h is described by its size Nh and, for every target variable Y, by its mean Mh and standard deviation Sh. The optimisation does not create new strata; it aggregates the atomic strata into a smaller number of final strata, so that the resulting design minimises the sample size required to satisfy the precision constraints.

Formally, let the atomic strata be indexed by h = 1, ..., L and let the target be an aggregation into K final strata. An assignment maps each atomic stratum to one of the K clusters. For any candidate aggregation, the minimum sample size that satisfies a set of coefficient-of-variation (CV) constraints, one per target variable, is computed by the Bethel-Chromy algorithm \citep{bethel1989,chromy1987}, which solves the underlying convex allocation problem and returns the optimal per-stratum sample sizes together with the achieved CVs. The algorithm computes a continuous allocation $n_k^*$, sends any stratum for which the computed sample size would exceed its population to a full census and re-solves on the remaining strata, iterating until no stratum is over-allocated, and takes the ceiling of the resulting continuous values component-wise to obtain the integer per-stratum sample sizes reported throughout this paper. This component-wise ceiling, $\tilde n_k=\lceil n_k^*\rceil$, is feasible but not guaranteed to be the smallest integer allocation meeting the precision constraints: the continuous optimum $C_B^{\mathrm{cont}}(z)=\sum_k n_k^*$ is a lower bound on, and generally strictly less than, the reported quantity $C_B^{\mathrm{reported}}(z)=\sum_k\lceil n_k^*\rceil$ that this paper calls the Bethel sample size throughout. Every method compared in this study is scored by the same $C_B^{\mathrm{reported}}$, so the comparison between methods remains valid, but ``minimum integer sample size'' would overstate what is actually computed. An atomic or final stratum of size one has a standard deviation of zero by construction, contributing no within-stratum variance; a final stratum that ends up with size one is sent directly to census, avoiding division by zero in the usual variance formula. Where later sections refer to a corrected Bethel-Chromy allocator, the correction is this census-and-iterate handling, applied consistently to every solver's decoded partition, not a change to the underlying convex allocation itself. The objective to be minimised is therefore the total Bethel sample size as a function of the aggregation, and the search space is the set of all assignments of L atomic strata to K clusters. Two features of this objective matter for everything that follows.

First, the objective is combinatorial: the number of possible aggregations grows like K$^L$, so exhaustive search is infeasible beyond very small L. This count is the number of labelled assignments, including those that leave one or more clusters empty; it therefore overstates the number of distinct unlabelled partitions, which, when every cluster is required to be non-empty, is given by the Stirling number $S(L,K)$ rather than $K^L/K!$. The order of growth is unaffected, but the exact count is not used anywhere in this paper beyond this qualitative statement. 

Second, the objective is not a simple pairwise function of the atomic strata; it is defined implicitly through the Bethel allocation, which couples all strata assigned to the same cluster. Any method that replaces this objective with a simpler surrogate, such as within-cluster homogeneity, is optimising a proxy rather than the sample size itself, a distinction that becomes central when comparing the approaches below.

\section{Classical solution: the genetic algorithm in SamplingStrata}
\label{sec:classical}
The reference solution is the genetic algorithm implemented in the R package SamplingStrata \citep{barcaroli2014,ballin2013}, which can be considered as one of the standard tools for this problem in official statistics. A genetic algorithm is a population-based stochastic optimiser inspired by natural selection. It maintains a population of candidate solutions, each encoded as a chromosome, and evolves the population across generations through selection, crossover, and mutation, retaining the fittest individuals at each step, in line with the general scheme implemented in standard genetic-algorithm software such as genalg \citep{willighagen2014}.

For the stratification problem the encoding is direct: a chromosome is a vector of length L whose hth entry is the cluster label assigned to atomic stratum h. The fitness of a chromosome is the total sample size returned by the Bethel-Chromy allocation for the aggregation it encodes; lower is fitter. Crucially, the genetic algorithm evaluates the true objective, the Bethel sample size, at every fitness evaluation, rather than a surrogate. It therefore searches directly in the space that matters, and its solutions define the practical benchmark against which every other method is measured. 

Formally, let $z=\left({z}_{1},\ldots ,{z}_{L}\right)$, with ${z}_{h}\in \{1,\ldots ,K\}$, denote a candidate aggregation, where ${z}_{h}=k$ means that atomic stratum $h$ is assigned to final stratum $k$. For each target variable ${Y}_{j}$, $j=1,\ldots ,J$, atomic stratum $h$ is characterised by its population size ${N}_{h}$, mean ${\overline{Y}}_{hj}$ and variance ${S}_{hj}^{2}$. The candidate aggregation determines the population size of final stratum $k$ as

$${N}_{k}\left(z\right)=\sum_{h=1}^{L} {N}_{h} I\left({z}_{h}=k\right),$$

and its mean for target variable ${Y}_{j}$ as

$${\overline{Y}}_{kj}\left(z\right)=\frac{\sum_{h=1}^{L} {N}_{h}{\overline{Y}}_{hj} I\left({z}_{h}=k\right)}{{N}_{k}\left(z\right)}.$$

The corresponding aggregated variance is

$${S}_{kj}^{2}\left(z\right)=\frac{\sum_{h:{z}_{h}=k} \left[\left({N}_{h}-1\right){S}_{hj}^{2}+{N}_{h}{\left\{{\overline{Y}}_{hj}-{\overline{Y}}_{kj}\left(z\right)\right\}}^{2}\right]}{{N}_{k}\left(z\right)-1}.$$

Conditional on this aggregation, let $n=\left({n}_{1},\ldots ,{n}_{K}\right)$ denote a candidate allocation. Under simple random sampling without replacement within each final stratum, the variance of the estimator of the population total of ${Y}_{j}$ is

$${V}_{j}\left(z,n\right)=\sum_{k=1}^{K} {N}_{k}^{2}\left(z\right)\left(1-\frac{{n}_{k}}{{N}_{k}\left(z\right)}\right)\frac{{S}_{kj}^{2}\left(z\right)}{{n}_{k}}.$$

If ${T}_{j}=\sum_{h=1}^{L} {N}_{h}{\overline{Y}}_{hj}$ is the population total of ${Y}_{j}$ and ${c}_{j}$ is its target coefficient of variation, the Bethel allocation problem associated with partition $z$ can be written as

$${C}_{B}\left(z\right)=\mathrm{min}_{n}\sum_{k=1}^{K} {n}_{k}$$

subject to

$$\frac{\sqrt{{V}_{j}\left(z,n\right)}}{\left|{T}_{j}\right|}\le {c}_{j},  j=1,\ldots ,J,$$

and

$${n}_{k}^{\mathrm{min}}\le {n}_{k}\le {N}_{k}\left(z\right),  k=1,\ldots ,K.$$

where ${n}_{k}^{\mathrm{min}}$ is the minimum admissible sample size in final stratum k, set in this study to the SamplingStrata default of 2, the minimum needed to estimate a within-stratum variance. The fitness assigned to chromosome $z$ is therefore the value ${C}_{B}\left(z\right)$ returned by the Bethel--Chromy allocation. For notational simplicity, throughout the remainder of this paper $C_B(z)$ denotes the reported, rounded allocation $C_B^{\mathrm{reported}}(z)$ introduced in Section~\ref{sec:problem} above, unless the continuous optimum is explicitly indicated; every sample size reported in this paper, including 105, 129 and 160, is a value of this reported quantity, not of the continuous optimum. The complete optimal-stratification problem solved by the genetic algorithm is

$${z}^{\star }=\mathrm{arg min}_{z\in \{1,\ldots ,K{\}}^{L}} {C}_{B}\left(z\right),$$

possibly subject to the additional requirement that all $K$ final strata be non-empty. Thus, each fitness evaluation contains an inner allocation problem: the chromosome determines the aggregation, whereas the Bethel--Chromy procedure determines the smallest feasible sample allocation for that aggregation. This nested structure also clarifies why the genetic algorithm evaluates the actual survey-design objective rather than a surrogate criterion.

SamplingStrata wraps this in a complete workflow: it builds the atomic strata from the frame, runs the genetic optimisation for a chosen K (or scans several values of K), applies the Bethel allocation, and returns the aggregation, the per-stratum allocation, and the achieved CVs. The genetic algorithm is stochastic, so in practice it is run for many generations and, where needed, restarted, to reduce the chance of settling in a poor local optimum. In the experiments below it provides the classical benchmark sample size.

\section{Quantum computing solution techniques}
\label{sec:quantum}
The true Bethel objective is not itself a quadratic function of the assignment variables. To obtain a problem compatible with quantum and quantum-inspired optimisers, the optimal-stratification problem is therefore reformulated by replacing the Bethel sample size with a population-weighted pairwise dispersion surrogate. Using binary one-hot indicators for the assignment of $L$ atomic strata to $K$ final strata, both the surrogate criterion and the assignment penalties are quadratic, yielding a quadratic unconstrained binary optimisation (QUBO) problem. QUBO is the natural input format for several quantum and quantum-inspired optimisers, in domains ranging from clustering to feature selection \citep{nembrini2021}. This section describes the surrogate formulation, its mapping onto three different quantum platforms, and, for each platform, how much of the resulting workflow is genuinely quantum and how much remains classical.

Let $z=\left({z}_{1},\ldots ,{z}_{L}\right)$, with ${z}_{h}\in \{1,\ldots ,K\}$, denote a candidate aggregation of the $L$ atomic strata into $K$ final strata. As in the Bethel formulation, ${z}_{h}=k$ means that atomic stratum $h$ is assigned to final stratum $k$. The same assignment can be represented through the binary indicators

$${x}_{hk}\mathrm{=I}\left({z}_{h}=k\right),  h=1,\ldots ,L, k=1,\ldots ,K,$$

subject to the one-hot constraints

$$\sum_{k=1}^{K} {x}_{hk}=1,  h=1,\ldots ,L.$$

Thus, the vector $z$ used in the Bethel formulation and the binary matrix $X=\left({x}_{hk}\right)$ used in the QUBO formulation represent the same discrete partition. In particular, ${x}_{hk}{x}_{h'k}=1$ if and only if atomic strata $h$ and $h'$ are both assigned to final stratum $k$.

\subsection{Pairwise dispersion surrogate}
For each atomic stratum $h$, define a feature vector ${u}_{h}$ containing the standardised means and standard deviations of the $J$ target variables:

$${u}_{h}=\left(\frac{{\overline{Y}}_{h1}-{\mu }_{{\overline{Y}}_{1}}}{{\sigma }_{{\overline{Y}}_{1}}},\ldots ,\frac{{\overline{Y}}_{hJ}-{\mu }_{{\overline{Y}}_{J}}}{{\sigma }_{{\overline{Y}}_{J}}},\frac{{S}_{h1}-{\mu }_{{S}_{1}}}{{\sigma }_{{S}_{1}}},\ldots ,\frac{{S}_{hJ}-{\mu }_{{S}_{J}}}{{\sigma }_{{S}_{J}}}\right).$$

Each $\mu$ and $\sigma$ is the unweighted mean and population standard deviation (denominator $L$, not $L-1$) of the corresponding column taken across the $L$ atomic strata; no weighting by $N_h$ is applied at this step, and a zero standard deviation is replaced by one, leaving the corresponding centred feature at zero for every atomic stratum. For the twenty-stratum case study, the values actually used are $\mu_{\overline{Y}_1}=191.59$, $\sigma_{\overline{Y}_1}=222.66$, $\mu_{\overline{Y}_2}=1317.50$, $\sigma_{\overline{Y}_2}=1175.18$, $\mu_{S_1}=38.03$, $\sigma_{S_1}=32.49$, $\mu_{S_2}=566.95$, $\sigma_{S_2}=634.74$.

The dissimilarity between atomic strata $h$ and $h'$ is then defined as

$${d}_{hh'}=\left\Vert{u}_{h}-{u}_{h'}\right\Vert,$$

and the corresponding population-weighted pairwise coefficient is

$${w}_{hh'}=\sqrt{{N}_{h}{N}_{h'}}\,{d}_{hh'}.$$

This weighting, plain Euclidean distance scaled by the square root of the product of populations, is a design choice for the surrogate, not a direct instance of the pairwise identity of Section~\ref{sec:why-surrogate}, which is stated in terms of the squared distance and the full product $N_h N_{h'}$; the two differ in both the power of the distance and the population weighting, and the relationship between them is heuristic rather than derived.

Other definitions of ${d}_{hh'}$ can be used without changing the QUBO structure. The essential requirement is that ${w}_{hh'}$ be fixed before the optimisation and therefore not depend on the unknown assignment.

A quadratic measure of within-stratum heterogeneity is

$${Q}_{\mathrm{disp}}\left(X\right)=\sum_{k=1}^{K} \sum_{1\le h<h'\le L} {w}_{hh'}{x}_{hk}{x}_{h'k}.$$

For a given pair $\left(h,h'\right)$, the term ${w}_{hh'}$ contributes to the objective only when the two atomic strata are assigned to the same final stratum. Minimising ${Q}_{\mathrm{disp}}$ therefore discourages the aggregation of atomic strata that are far apart in the standardised feature space.

\subsection{Incorporating the assignment constraints}
\label{sec:constraints}
The one-hot constraints can be incorporated into the objective by means of a quadratic penalty, $Q_{\mathrm{pen}}(X)=\lambda\sum_h(1-\sum_k x_{hk})^2$, where $\lambda>0$ (a different symbol from the annealing schedule function $A(t)$ of Section~\ref{sec:dwave}). Expanding this penalty using $x_{hk}^2=x_{hk}$ gives the complete QUBO objective as the quadratic polynomial

$$Q\left(X\right)=-\lambda\sum_{h=1}^{L} \sum_{k=1}^{K} {x}_{hk}+\sum_{k=1}^{K} \sum_{1\le h<h'\le L} {w}_{hh'}{x}_{hk}{x}_{h'k}+2\lambda\sum_{h=1}^{L} \sum_{1\le k\mathrm{<l\le }K} {x}_{hk}{x}_{hl}+\lambda L,$$

with the full expansion given in Appendix~\ref{app:math-penalty}. Throughout this paper, quadratic QUBO coefficients are stated as a single weight per unordered pair of variables (equivalently, an upper-triangular coefficient-matrix convention, no factor of two, no lower-triangular duplicate); this convention is used consistently for every reported energy value, including 501.0, 566.0 and 936.61.

The coefficient $\lambda$ must be large enough to make violations of the one-hot constraints unattractive, but an excessively large value can compress the differences among feasible solutions and worsen numerical conditioning. Throughout this study $\lambda=3\max_{h,h'}w_{hh'}=476.92$; a simple sufficient condition guaranteeing that no violation is ever profitable is $\lambda > \max_h \sum_{h'\neq h} w_{hh'}$, the largest total pairwise weight incident on any single atomic stratum. For the twenty-stratum case study this quantity is 1{,}729.23, so the penalty actually used does \emph{not} meet this sufficient condition; the smaller value was retained because it kept the QUBO better conditioned, and every decoded solution reported in this paper was checked and, where necessary, repaired for one-hot violations rather than relying on the penalty to rule them out by construction, as described for each solver in the sections that follow. The repair rule is the same throughout: an atomic stratum assigned to more than one cluster is kept in whichever of its assigned clusters yields the smallest repaired dispersion value, all its other assignments being cleared; a stratum assigned to no cluster is placed in whichever cluster yields the smallest repaired dispersion value; and if a final cluster is left empty after this step, the atomic stratum, drawn only from clusters that would still retain at least one other atomic stratum after its removal, whose reassignment to the empty cluster increases the surrogate least is moved there, so that this step cannot itself create a new empty cluster. The common QUBO energy and the downstream Bethel sample size are always recomputed on the repaired, feasible partition, never on the raw decoded bits. All solver results reported in Table~\ref{tab:results} and Figure~\ref{fig:results} were feasible after repair and contained exactly four non-empty final strata.

For a sufficiently large coefficient, the penalty above enforces one-hot assignment. With the smaller coefficient used here, as just discussed, feasibility was instead verified and restored through the repair procedure described above, rather than guaranteed by the penalty alone. Neither the penalty nor the repair rule as stated ensures that every one of the $K$ final strata is non-empty on its own; the additional repair step above handles that case separately. If non-empty final strata are required, additional constraints, penalty terms, or a classical repair procedure must be introduced.

\subsection{Relationship with the Bethel objective}
\label{sec:relationship}
The Bethel objective $C_B(z)$, defined in Section~\ref{sec:classical}, is not a quadratic polynomial in the assignment indicators: the mean of $Y_j$ in final stratum $k$, $\overline{Y}_{kj}(X)=\sum_h N_h\overline{Y}_{hj}x_{hk}/\sum_h N_h x_{hk}$, has an assignment-dependent denominator, the aggregated variances share this dependency, and the Bethel--Chromy allocation itself adds an inner iterative optimisation, so $C_B(z)$ has no direct QUBO representation. The connection to the quadratic surrogate can be seen from the pairwise identity

$$\sum_{h:{z}_{h}=k} {N}_{h}{\left({\overline{Y}}_{hj}-{\overline{Y}}_{kj}\right)}^{2}=\frac{1}{2{N}_{k}\left(z\right)}\sum_{h=1}^{L} \sum_{h'=1}^{L} {N}_{h}{N}_{h'}{\left({\overline{Y}}_{hj}-{\overline{Y}}_{h'j}\right)}^{2}{x}_{hk}{x}_{h'k}$$

(derivation in Appendix~\ref{app:math-pairwise}): for fixed $N_k$ the right-hand numerator is quadratic in the assignment indicators, but $N_k(z)$ itself varies with the partition, and dropping the factor $1/\{2N_k(z)\}$ to obtain a valid QUBO changes the criterion and can alter the ranking of partitions, especially when final-stratum sizes differ substantially.

Accordingly, the QUBO problem actually solved, $X^Q=\mathrm{arg\,min}_X Q(X)$, differs from the optimal-stratification problem of substantive interest, $z^B=\mathrm{arg\,min}_z C_B(z)$: the partition $X^Q$ must be decoded into $z^Q$ and evaluated classically through $C_B(z^Q)$. This is a general feature of every QUBO-based workflow below: each solves a within-cluster dispersion surrogate, a proxy for the Bethel sample size rather than the sample size itself, and the Bethel allocation is always computed classically afterwards, on the classical side.

\subsection{Two notions used throughout: the Hamiltonian and the Ising model}
\label{sec:notions}
Two terms recur in every quantum formulation and are worth defining once. In physics, the Hamiltonian of a system assigns an energy to each of its possible configurations; a physical system, left to itself and cooled, tends to settle into its lowest-energy, ground-state configuration. Here, the problem Hamiltonian is simply the operator representation of the QUBO surrogate, $H_P(X)=Q(X)$ as defined above, so that each binary assignment matrix corresponds to a physical configuration and the QUBO value to its energy; the ground state, $X^{\mathrm{GS}}=\mathrm{arg\,min}_X H_P(X)$, therefore corresponds to a global minimiser of the dispersion surrogate, not necessarily of the Bethel objective, and practical algorithms generally return low-energy samples rather than a certified ground state.

The Ising representation expresses the same energy through spin variables $s_i\in\{-1,+1\}$ rather than binary variables $x_i\in\{0,1\}$, related by $s_i=1-2x_i$; this substitution changes the coefficients and may add an irrelevant constant, but preserves the ordering of configurations and the set of minimisers (full derivation in Appendix~\ref{app:math-ising}). Throughout this paper, therefore, QUBO objective, Ising energy and problem Hamiltonian refer to equivalent representations of the same penalised dispersion surrogate, never to the Bethel sample-size objective itself.

\subsection{D-Wave QPU (quantum annealing)}
\label{sec:dwave}
D-Wave systems are quantum annealers \citep{kadowaki1998,johnson2011}. The QUBO/Ising formulation derived above already defines the optimisation problem independently of the solver; running it on a D-Wave QPU translates the same objective into the required coefficient representation, preserving feasible assignments, their relative ordering and the minimisers, apart from irrelevant additive constants or positive scaling factors. The search begins from the ground state of a simple driver Hamiltonian $H_0$ and evolves under $H(t)=A(t)H_0+B(t)H_P$, $0\le t\le T$, with the schedule functions $A(t)$ and $B(t)$ gradually reducing the influence of $H_0$ and increasing that of the problem Hamiltonian $H_P$ (a different use of $A$ from the penalty coefficient $\lambda$ of Section~\ref{sec:quantum}), so the system is driven from an easily prepared state towards low-energy configurations; under ideal, sufficiently slow evolution the system remains close to the instantaneous ground state, though on real hardware finite annealing time, noise and control errors mean the measured configuration need not be exact (full schedule details in Appendix~\ref{app:math-dwave-schedule}).

A further implementation step is required because the logical Ising model is dense, whereas the QPU has a fixed sparse connectivity graph. Minor embedding represents a logical variable by a chain of connected physical qubits, so the number of physical qubits can be substantially larger than the number of logical variables. This hardware-level transformation does not change the definition of the optimisation problem, but it can affect the quality of the search through chain length, chain strength, broken chains, coefficient scaling and hardware noise. The distinction between logical variables and physical qubits after embedding is therefore retained in the scaling analysis below.

Figure 1 shows how this qubit requirement grows with the number of atomic strata for both the one-hot and binary-label encodings, together with an illustrative estimate of the physical qubits needed after embedding onto the named hardware topology, against the qubit budget of that machine.

Quantum versus classical stages. In the pure QPU mode the annealing itself, the search for a low-energy configuration, is genuinely quantum. However, the construction of the QUBO, the embedding of its dense connectivity onto the fixed sparse qubit topology of the hardware (with chains of physical qubits representing one logical variable), the choice of penalty strength, the decoding of the measured bitstring, the repair of any one-hot violations, and the entire Bethel evaluation are classical. In the hybrid mode offered on the D-Wave cloud, a classical solver additionally decomposes the problem and calls the QPU only on sub-problems, so the quantum share is smaller still. In this study the QPU/hybrid modes were not accessible on the available plan, and the QUBO was solved with classical simulated annealing, which is entirely classical but exercises exactly the same problem formulation and decoding pipeline.

\begin{figure}[htbp]
\centering
\includegraphics[width=0.92\textwidth]{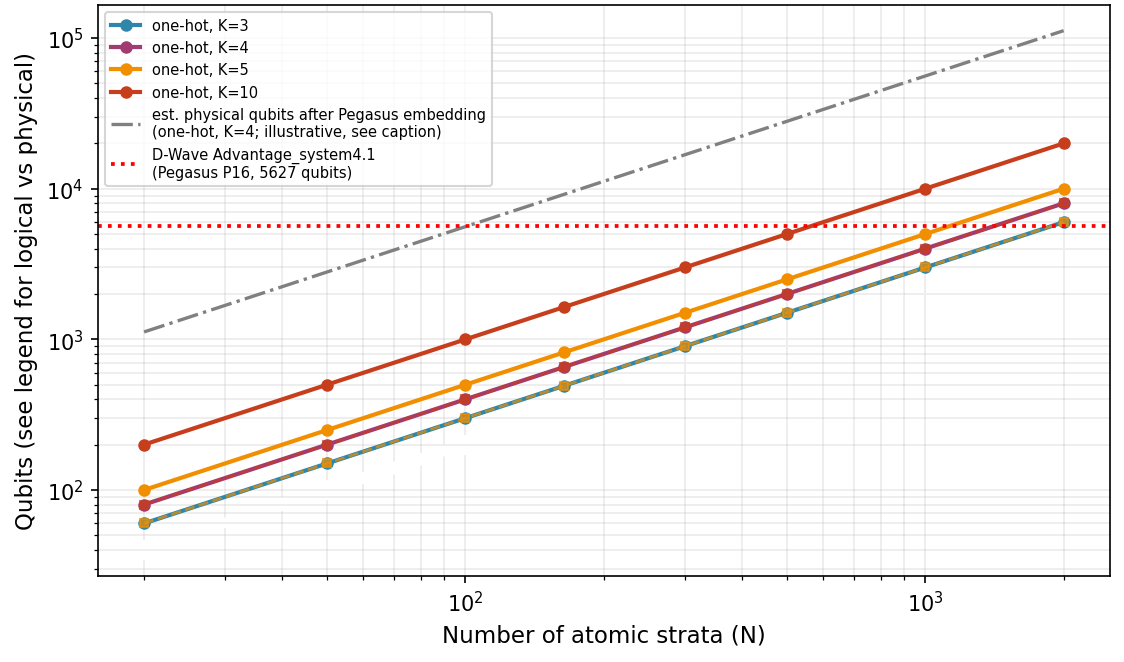}
\caption{D-Wave qubit scaling. Solid: logical variables under one-hot encoding. Dashed: logical variables under a binary-label encoding of the cluster index (ancilla variables needed to reduce the resulting higher-order same-cluster terms to quadratic form are not counted). Dash-dot: an illustrative, complete-graph-based estimate of physical qubits after minor-embedding, calibrated to the documented capacity of embedding a fully connected graph of about 150 variables onto the Pegasus P16 topology with chains of roughly 14 physical qubits each. The actual one-hot QUBO graph is markedly sparser than complete (about 28\% density at L = 20, K = 4, as detailed in Section~\ref{sec:dwave-paid}), so this curve is a conservative reference rather than a prediction of the true embedding requirement, which was not computed in this study. The dotted line marks the total qubit count of D-Wave Advantage\_system4.1, subsequently renamed Advantage\_system4 \citep{dwave2026solver} (Pegasus P16, 5,627 qubits), the machine referenced throughout this study; the comparison against this total is illustrative only; the number of variables a clique-based problem can actually reach is set by the topology's embedding capacity (about 150 variables, the K150 figure discussed in the text), not by the total qubit count directly, and the dash-dot curve should be read against that capacity rather than against the dotted line.}
\label{fig:dwave}
\end{figure}

\FloatBarrier
\subsection{IBM Quantum (gate-based QAOA)}
\label{sec:ibm}
IBM Quantum provides gate-based superconducting quantum processors. In this study, the stratification surrogate is addressed through the Quantum Approximate Optimisation Algorithm (QAOA), a hybrid variational method that alternates a parameterised cost-Hamiltonian evolution with a mixing evolution over $p$ circuit layers, the parameters being updated classically between measurements \citep{farhi2014}. The classical optimisation was carried out with COBYLA, a derivative-free method comparatively robust to finite-shot sampling noise; the exact layer count $p$, shot count and the ibm\_kingston backend details are given in Appendix~\ref{app:math-qaoa-params}.

Unlike the D-Wave, Dirac-3 and Amplify AE implementations, IBM does not use the one-hot representation with L$\times$K binary variables. Instead, each atomic stratum is assigned a binary label identifying its final stratum, requiring $b=\lceil {\mathrm{log}}_{2}K\rceil$ qubits per stratum, so a problem with L atomic strata requires Lb qubits; for K = 4, two qubits per stratum give forty qubits for the twenty-stratum problem.

This representation expresses the same underlying dispersion criterion as the other solvers, but through a different polynomial encoding. Determining whether two binary labels are equal requires considering their bits jointly: for K = 4, with qubits $(Z_{h1},Z_{h2})$ encoding the label of stratum $h$, the equality indicator is
$$\delta\left(z_h,z_{h'}\right)=\frac{1+Z_{h1}Z_{h'1}}{2}\cdot\frac{1+Z_{h2}Z_{h'2}}{2}=\frac{1}{4}\left(1+Z_{h1}Z_{h'1}+Z_{h2}Z_{h'2}+Z_{h1}Z_{h'1}Z_{h2}Z_{h'2}\right),$$
giving the cost Hamiltonian $H_C=\sum_{h<h'} w_{hh'}\,\delta\left(z_h,z_{h'}\right)$, the same pairwise weights $w_{hh'}$ used throughout this paper up to an identity term that does not affect the minimisers. This directly confirms why $\binom{20}{2}=190$ atomic-stratum pairs yield $2\times190=380$ two-body terms and $190$ four-body terms at K = 4, matching the counts reported for the real run below; IBM's Hamiltonian encodes the same dispersion criterion as the one-hot QUBO without being the same polynomial term by term, which is why the two solvers' raw Hamiltonian energies are not compared directly anywhere in this paper. Dropping the identity term from $H_C$ gives the operator actually implemented, $H_C^{\mathrm{impl}}$; a short exact calculation, given in Appendix~\ref{app:math-ibm-energy}, shows that its expectation under a uniform distribution over the four labels is precisely zero. This is the reason the near-zero cost energy measured on real hardware, discussed below, is exactly what the implemented Hamiltonian predicts for a broad, near-uniform output, not an anomaly requiring a separate offset or scaling explanation.

Once the variational parameters have been optimised, the final circuit is sampled repeatedly; each observed bit string decodes into a candidate partition, ranked by the encoded surrogate and then evaluated classically through the Bethel--Chromy allocation, exactly as in the other workflows.

A major practical limitation arises during transpilation, the compilation step that converts the abstract QAOA circuit into gates supported by the physical processor. IBM superconducting devices use a sparse heavy-hex connectivity graph, whereas the stratification Hamiltonian contains interactions between many pairs of atomic strata; when two qubits that must interact are not physically adjacent, the transpiler inserts SWAP gates to move their quantum states across the device. This routing process can increase circuit depth and gate count substantially, and since two-qubit gates are among the main sources of error on current hardware, the accumulated overhead can destroy the useful signal before the circuit is measured. For the twenty-stratum, K = 4 problem considered here, this connectivity constraint proved more restrictive than the number of available qubits.

Quantum versus classical stages. The preparation and measurement of the parameterised quantum state are quantum. The construction of the Hamiltonian, the transpilation of the circuit, the optimisation of the variational parameters, the decoding of the measured bit strings and the Bethel--Chromy evaluation are classical. QAOA should therefore be interpreted as a hybrid search procedure in which the quantum processor supplies samples from a parameterised distribution over candidate assignments, while the surrounding optimisation and the evaluation of the true survey-design objective remain classical.

Figure 2 illustrates the scaling of the IBM implementation. Panel (a) reports the number of qubits required by the binary-label encoding. Panel (b) reports the number of interaction terms in the resulting Hamiltonian. These counts describe the abstract representation of the problem; the number of physical two-qubit gates after transpilation and SWAP routing can be substantially larger.

\begin{figure}[htbp]
\centering
\includegraphics[width=0.92\textwidth]{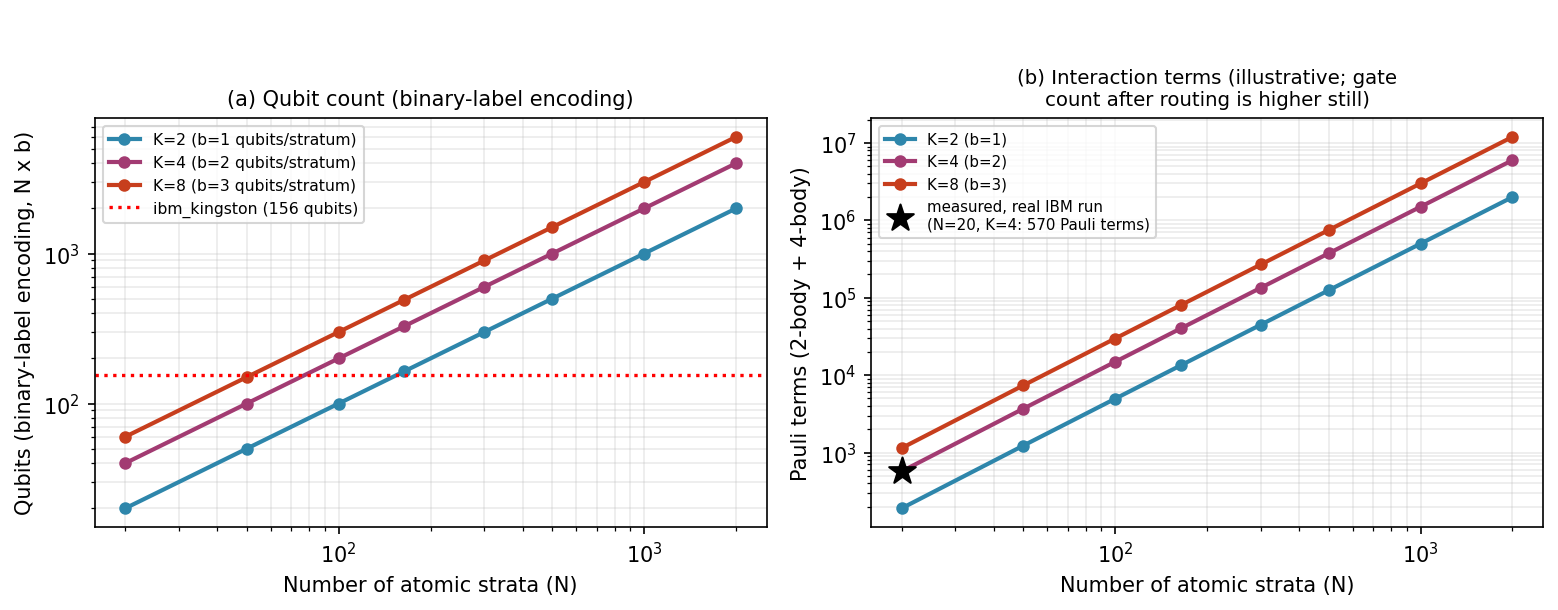}
\caption{IBM Quantum QAOA scaling under the binary-label encoding used in this study. Panel (a) shows the number of qubits required as the number of atomic strata increases. Panel (b) shows the corresponding number of two-body and four-body interaction terms. These are counts for the abstract Hamiltonian; the number of physical two-qubit gates after transpilation and SWAP routing is substantially larger. The curves for K are not powers of two (e.g.\ K = 3) do not account for the $2^b-K$ unused codewords per stratum, which would require additional penalty or ancilla terms to exclude; the K = 4 case used in this study's real run is unaffected, since it is an exact power of two.}
\label{fig:ibm}
\end{figure}

\subsection{QCi Dirac-3 (entropy quantum computing)}
\label{sec:dirac}
The QCi Dirac-3 is a photonic entropy-quantum-optimisation device. Unlike superconducting platforms, it provides all-to-all connectivity and supports both binary variables and multilevel variables, or qudits. It therefore does not require minor embedding onto a sparse hardware graph and does not incur the SWAP-routing overhead encountered in gate-based implementations.

For the present stratification problem, two representations are relevant.

The first is the same one-hot QUBO used for D-Wave and Amplify AE. Each atomic stratum is represented by K binary variables, with a one-hot constraint ensuring that it is assigned to exactly one final stratum. The population-weighted pairwise dispersion surrogate and the assignment penalties are then expressed as a quadratic polynomial.

The second possibility is to represent the final-stratum label of each atomic stratum by a single K-level variable. This integer or qudit encoding is more compact, since it requires one decision variable per atomic stratum rather than K binary variables. However, compactness alone does not guarantee a suitable objective function.

An initially considered formulation penalised pairs of labels through the squared difference $(x_h-x_{h'})^2$, but this is inappropriate for nominal cluster labels: it introduces an artificial ordering (labels 0 and 3 are treated as more different than 0 and 1, though both simply indicate different clusters), and with non-negative weights it is minimised by assigning every atomic stratum the same label, the degenerate single-cluster solution, irrespective of the intended number of final strata. These problems arise from the squared-difference comparison, not from the qudit representation itself; a valid label-based formulation would need to distinguish only equal from unequal labels, independently of numerical value, but expressing such an equality indicator as a low-degree polynomial generally requires higher-order terms or additional auxiliary variables, reducing the compact encoding's advantage.

For this reason, the present study uses the one-hot QUBO. It is less compact, but it provides a direct quadratic and label-invariant representation of the selected dispersion surrogate and makes the Dirac-3 result directly comparable with D-Wave and Amplify AE.

The Dirac-3 optimises the supplied polynomial directly on its photonic hardware. Its all-to-all connectivity avoids the embedding chains required by quantum annealers and the SWAP gates required by sparse gate-based processors. It also does not use a variational quantum circuit with a classical parameter-optimisation loop.

Once the device has returned candidate assignments, the remaining steps are classical. The solutions are decoded, any one-hot violations are repaired, and each candidate partition is evaluated through the Bethel--Chromy allocation. As in the other workflows, the device minimises the dispersion surrogate rather than the Bethel sample size itself.

Quantum versus classical stages. According to the vendor's description, the optimisation of the submitted polynomial is performed on photonic hardware. The construction and scaling of the QUBO, the decoding and repair of the returned solution, and the final Bethel--Chromy evaluation remain classical. The underlying physical mechanism was not independently verified in this study and is therefore reported as described by the provider.

Figure 3 summarises the scaling of the two representations. The one-hot formulation requires L$\times$K binary variables, whereas the integer or qudit formulation would require only L variables. Under the free-tier limit used in this study, the number of one-hot decision variables is the binding constraint and determines the maximum problem size that can be submitted.

\begin{figure}[htbp]
\centering
\includegraphics[width=0.92\textwidth]{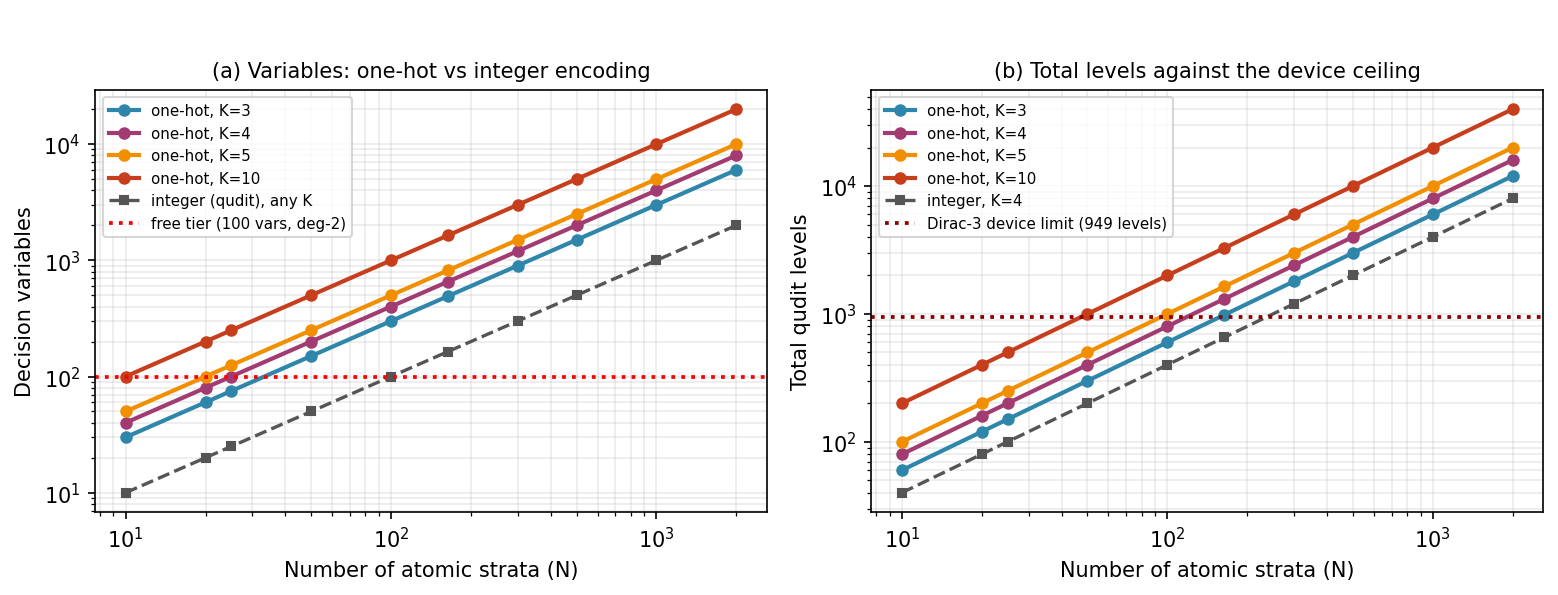}
\caption{QCi Dirac-3 scaling. Panel (a) compares the number of decision variables required by the one-hot encoding, $L\times K$, and by the integer or qudit encoding, $L$. Panel (b) reports the total number of qudit levels required. The free-tier decision-variable limit is binding for the one-hot formulation used in this study.}
\label{fig:dirac}
\end{figure}

\FloatBarrier
\subsection{Fixstars Amplify AE (GPU Ising machine)}
\label{sec:amplify}
The fourth solver is not a quantum computer at all, and it is included precisely for that reason. The Fixstars Amplify Annealing Engine \citep{fixstars2026} is a GPU-based Ising machine: it minimises the same one-hot QUBO Hamiltonian used for D-Wave and the Dirac-3 (the gate-based encoding used for IBM differs, as noted in Section~\ref{sec:ibm}), using annealing-family heuristics executed on graphics processors rather than on qubits. The engine itself scales to large problems, but the free Basic plan used in this study, intended for evaluation and testing, is capped at 8,192 bits for a fully-connected graph; our 80-variable one-hot model sits far below either limit, and it is offered free of charge, so it can be applied to the present problem without any of the quota restrictions that constrained the other three platforms.

Its role in this study is that of a control. Because it accepts a densely connected QUBO directly, there is no minor-embedding, no chain of physical qubits per logical variable, no transpilation onto a sparse lattice and no SWAP overhead; and because it is a classical machine, there is no gate noise and no decoherence. It therefore removes, in a single stroke, the principal quantum-specific hardware limitations considered here, while leaving the mathematical problem exactly as it was, though its own native constraint handling and finite search budget mean it is not literally solving the identical penalised polynomial explored by the other solvers, so a shortfall here cannot be blamed on quantum-specific hardware effects, though it cannot rule out other algorithmic or computational limitations.

The formulation handed to it is the same feasible objective and assignment constraints used for D-Wave and the Dirac-3, represented through the solver's native constraint mechanism: the same one-hot encoding with L $\times$ K binary variables, the same population-weighted intra-cluster distance objective, and the same one-hot penalty weight, set to three times the largest objective coefficient. The one implementation difference is that the Amplify SDK expresses the one-hot condition as a first-class constraint object rather than requiring the penalty to be folded into the QUBO matrix by hand; the two representations are equivalent on the feasible set, but the solver is not literally handed the same penalised polynomial, so it is not necessarily exploring the same energy landscape away from that set.

Quantum versus classical stages. None of it is quantum. Amplify AE is a quantum-inspired solver: it borrows the Ising formulation and the annealing metaphor from quantum annealing, but it runs entirely on classical GPU hardware and makes no use of quantum mechanical effects. Including it alongside the three quantum platforms is what makes the comparison in Section~\ref{sec:comparative} interpretable, because it establishes what the QUBO formulation can deliver when quantum-specific hardware limitations are absent and a strong classical heuristic is used.

\begin{figure}[htbp]
\centering
\includegraphics[width=0.92\textwidth]{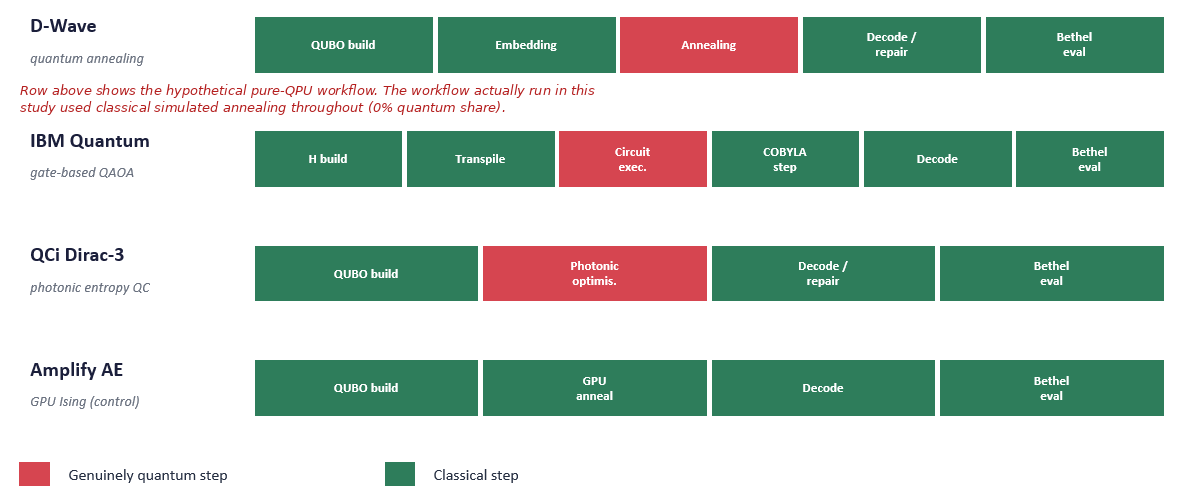}
\caption{How each solver's workflow divides into computational stages. Each bar allocates equal width to each named stage of that solver's own pipeline as a display convention; segment width does not represent measured runtime, computational cost, or any other quantified share of ``how quantum'' the pipeline is. What differs across solvers is the number of stages classified as genuinely quantum (0 of 4 for Amplify AE, 1 of 5 for D-Wave, 1 of 6 for IBM, 1 of 4 for Dirac-3), not a measured proportion of effort. The D-Wave row shows the hypothetical pure-QPU workflow for reference; the workflow actually run in this study, noted directly on the figure, used classical simulated annealing throughout and has no genuinely quantum stage.}
\label{fig:qshare}
\end{figure}

\FloatBarrier
\section{Case study: the reduced swissmunicipalities dataset}
\label{sec:case-study}
The experiments use the swissmunicipalities dataset, a standard benchmark distributed with the SamplingStrata package. It records, for each Swiss municipality, demographic and land-use variables. The analysis is restricted to the 589 municipalities of a single region (REG = 1 in the dataset's own coding) and to two auxiliary variables, total population (POPTOT) and total area (HApoly), which are discretised to form the atomic strata, and two target variables, the built-up area (Airbat) and the wooded area (Surfacesbois), on which the precision constraints are imposed.

Each auxiliary variable is discretised into five classes using the categorisation distributed with the dataset (POPTOT.cat, HApoly.cat); the classes are unequal in width and in the number of municipalities they contain, reflecting the right-skewed distribution of both variables rather than equal-width or equal-frequency binning. For POPTOT the five classes span, respectively, 27--1{,}601 (450 municipalities), 1{,}644--5{,}182 (89), 5{,}341--12{,}141 (32), 13{,}933--29{,}559 (16) and 124{,}914--177{,}964 (2) inhabitants; for HApoly the corresponding ranges are 32--1{,}062 (416 municipalities), 1{,}100--3{,}031 (107), 3{,}144--6{,}723 (44), 7{,}093--11{,}907 (17) and 16{,}503--28{,}225 (5) hectares. The two largest classes of each variable are populated by a handful of outlying municipalities (the largest Swiss cities and the largest, mostly Alpine, municipalities by area), which is why several of the resulting atomic strata are singletons.

The precision requirement is a coefficient of variation of five percent on each target variable, and the aggregation target is K = 4 final strata. The two auxiliary variables are each categorised into five classes, so the full cross-classification yields up to twenty-five atomic strata; on the actual data twenty of these are non-empty, four of which are singletons. This deliberately coarse frame of twenty atomic strata is the common input on which every method is run. Table~\ref{tab:dataset} gives the population, mean and standard deviation of both target variables for each of the twenty atomic strata.

\begin{table}[htbp]
\centering
\small
\begin{tabular}{@{}cc r rr rr@{}}
\toprule
\textbf{POPTOT.cat} & \textbf{HApoly.cat} & \textbf{$N_h$} & \textbf{$\overline{Y}_{h1}$ (Airbat)} & \textbf{$S_{h1}$} & \textbf{$\overline{Y}_{h2}$ (Surfacesbois)} & \textbf{$S_{h2}$} \\
\midrule
1 & 1 & 337 & 15.4 & 11.1 & 123.2 & 122.1 \\
1 & 2 & 72 & 28.4 & 19.3 & 700.3 & 304.6 \\
1 & 3 & 30 & 35.4 & 26.0 & 1138.2 & 586.1 \\
1 & 4 & 9 & 22.9 & 20.2 & 1175.9 & 600.5 \\
1 & 5 & 2 & 47.5 & 53.0 & 1458.0 & 1514.6 \\
2 & 1 & 52 & 70.5 & 34.8 & 82.2 & 87.0 \\
2 & 2 & 24 & 87.7 & 35.7 & 664.2 & 418.8 \\
2 & 3 & 9 & 98.3 & 34.5 & 1909.8 & 806.4 \\
2 & 4 & 3 & 122.7 & 46.8 & 4123.3 & 2318.5 \\
2 & 5 & 1 & 118.0 & 0.0 & 4163.0 & 0.0 \\
3 & 1 & 17 & 145.9 & 61.0 & 85.5 & 100.6 \\
3 & 2 & 5 & 116.2 & 27.9 & 631.0 & 366.0 \\
3 & 3 & 3 & 181.7 & 110.4 & 1831.0 & 810.1 \\
3 & 4 & 5 & 180.6 & 48.2 & 1970.8 & 967.8 \\
3 & 5 & 2 & 172.5 & 120.9 & 2281.5 & 1781.2 \\
4 & 1 & 10 & 169.3 & 55.2 & 53.6 & 44.3 \\
4 & 2 & 5 & 238.8 & 55.7 & 615.6 & 510.5 \\
4 & 3 & 1 & 351.0 & 0.0 & 1641.0 & 0.0 \\
5 & 2 & 1 & 773.0 & 0.0 & 67.0 & 0.0 \\
5 & 3 & 1 & 856.0 & 0.0 & 1635.0 & 0.0 \\
\bottomrule
\end{tabular}
\caption{The twenty non-empty atomic strata of the case study, indexed by the POPTOT and HApoly discretisation classes. $N_h$ is the number of municipalities in the stratum; $\overline{Y}_{h1}, S_{h1}$ and $\overline{Y}_{h2}, S_{h2}$ are the mean and standard deviation of Airbat and Surfacesbois respectively. A standard deviation of zero indicates a singleton stratum.}
\label{tab:dataset}
\end{table}
\FloatBarrier

\subsection{Why the study is limited to twenty atomic strata}
\label{sec:why20}
The choice of a twenty-stratum frame is not statistical but operational: it is the largest problem that all three quantum platforms can accept on their freely available tiers, so that the same frame can be run on every platform without pre-grouping. Each platform imposes a different limit, and the binding one is the tightest across the three.

On the IBM gate-based platform the constraint is circuit depth rather than qubit count. A direct binary-label encoding of the twenty-stratum problem at K = 4 (two qubits per stratum, as described in Section~\ref{sec:ibm}) needs forty qubits, which the hardware has in abundance, but the transpiled circuit reaches a depth of several thousand two-qubit gates, already at the edge of what current noisy processors can execute before the signal is lost. Larger frames would further increase circuit depth and would be expected to make noise-dominated output even more likely on hardware of comparable connectivity and fidelity.

On the QCi Dirac-3 the free tier caps a degree-two problem at one hundred decision variables. This limit value, 100, is stated directly and unambiguously by the device API itself: an earlier submission, at an exploratory 164-atomic-stratum frame tried before the twenty-stratum frame was settled on, was rejected with the response ``Number of variables `164' in problem is greater than the free-tier device limit `100' for polynomial with degree `2''' \citep{qci2026}. That earlier attempt used the integer (qudit) encoding described in Section~\ref{sec:dirac}, one decision variable per atomic stratum, which is why the count in the API response is 164 rather than the 656 that a one-hot encoding of the same frame would require at K = 4; the rejection was itself one of the reasons the frame was subsequently reduced. The one-hot QUBO used throughout the rest of this study needs L $\times$ K variables, so at K = 4 the ceiling of one hundred variables would be reached at L = 25 atomic strata under this reading; twenty strata (eighty variables) sit under it and was, in practice, accepted. A finer frame approaching or exceeding L = 25 would be expected to risk rejection, though this was not tested directly at the boundary.

On the D-Wave platform the classical simulated-annealing solver used here has no such hard ceiling, but to keep the comparison meaningful it is run on the identical twenty-stratum frame. Taken together, the three limits make twenty atomic strata the natural common ground: fine enough to be a genuine optimisation problem, coarse enough to run on every platform without approximation or pre-grouping.

The frame, the target variables, the precision constraint and the final Bethel-Chromy evaluation are common to all four solvers; the computational budget spent by each is not, because it is itself constrained by what each free tier allows. Table~\ref{tab:budgets} makes this explicit, so that the results in Section~\ref{sec:comparative} are read as a comparison under each platform's actual free-tier constraints, not as a comparison with a uniform compute budget.

\begin{table}[htbp]
\centering
\small
\begin{tabular}{@{}p{2.4cm}p{4.6cm}p{5.0cm}@{}}
\toprule
\textbf{Solver} & \textbf{Budget actually used} & \textbf{Binding constraint} \\
\midrule
Fixstars Amplify AE & Single call, approximately five seconds & Free Basic plan timeout, not variable count \\
D-Wave (classical SA) & Best-of-ten quick search, fewer sweeps and restarts than the independent verification search & QPU not authorised on the free plan; SA run kept short for comparability \\
IBM Quantum & Two COBYLA iterations, each queued as a separate job & Ten minutes of QPU time per rolling twenty-eight-day window; no session mode on the free plan \\
QCi Dirac-3 & Single submission, ten returned solutions & Free-tier cap of 100 decision variables at polynomial degree two \\
\bottomrule
\end{tabular}
\caption{Computational budget actually used by each solver on the common twenty-stratum frame, and the free-tier constraint that determined it. Exact sweep and restart counts for the D-Wave quick search were not systematically logged and are reported qualitatively.}
\label{tab:budgets}
\end{table}

The qubit and variable scaling implied by each encoding, together with the figure illustrating it, is given alongside the description of the corresponding platform: Section~\ref{sec:dwave} for D-Wave, Section~\ref{sec:ibm} for IBM, and Section~\ref{sec:dirac} for the Dirac-3.

\FloatBarrier
\section{Comparative analysis of the results}
\label{sec:comparative}
Every method was run on the identical twenty-stratum frame, with the same two target variables, the same five-percent precision constraint, K = 4, and the same corrected Bethel-Chromy allocator used for evaluation. The classical benchmark, the best result obtained from the SamplingStrata genetic algorithm operating directly on the atomic strata, achieves a total sample size of 129; the genetic algorithm is stochastic, and this figure should be read as the best-known result obtained rather than a certified optimum, since restart and seed distributions were not systematically logged in this study. As a reference point, the no-aggregation baseline, an optimised Bethel allocation applied directly to all twenty atomic strata without any clustering step, requires 105; because the starting frame is already coarse, aggregating to four strata sacrifices some efficiency, so this baseline is smaller than the best aggregated result found.

The meaningful question is therefore how close each aggregation method comes to the best classical aggregator result at K = 4, namely 129. The two classical references bracket the picture from opposite sides. The genetic algorithm, which optimises the Bethel sample size directly, reaches the benchmark of 129. The KmeansSolution routine requires 312: it clusters the standardised atomic-stratum features to produce a starting point for the genetic search, using the same feature construction as the QUBO surrogate but with a classical k-means step in place of any subsequent optimisation, and on its own, without the subsequent genetic optimisation, it is far from optimal. It is included in Table~\ref{tab:results} as a reference initialisation, not as a solver competing on the same footing as the four QUBO-based methods and the genetic algorithm; its exact configuration, in particular the number of k-means starts, is inherited from SamplingStrata's default and was not varied in this study.

\begin{table}[htbp]
\centering
\small
\begin{tabular}{@{}lrrp{5.2cm}@{}}
\toprule
\textbf{Method} & \textbf{n} & \textbf{Gap vs 129} & \textbf{Nature of the result} \\
\midrule
Baseline (20 atomic strata) & 105 & - & no aggregation; fine granularity \\
SamplingStrata GA (benchmark) & 129 & 0\% & best GA result obtained \\
KmeansSolution, K = 4 & 312 & +142\% & GA initialisation, not a competing solver \\
Fixstars Amplify AE (GPU Ising) & 160 & +24\% & best-known QUBO solution found \\
D-Wave formulation, classical SA & 162 & +26\% & a weaker local optimum of the same QUBO \\
IBM QAOA, K = 4 (40 qubits, real HW) & 262 & +103\% & severely truncated run; consistent with strong noise and under-optimisation \\
QCi Dirac-3 QUBO, K = 4 (real HW) & 286 & +122\% & settled far from the best-known optimum \\
\bottomrule
\end{tabular}
\caption{Sample size by method on the common twenty-stratum frame (K = 4, CV = 5\%).}
\label{tab:results}
\end{table}
\FloatBarrier

The four surrogate-based solvers then separate into two groups of clearly different quality. Fixstars Amplify AE, the GPU Ising machine, reaches an objective value of 501.0 in about five seconds, giving a sample size of 160; an independent, more thorough simulated-annealing search run as a check converges repeatedly on that same value, so 501.0 is treated throughout as the best-known solution of this QUBO, though without a certificate of global optimality. D-Wave's own quick best-of-ten simulated-annealing search, run with a smaller number of sweeps and restarts, settles instead on a visibly weaker local optimum, implying an objective of about 566.0, and correspondingly returns a sample size of 162: close to, but measurably above, the best-known value. That a somewhat worse QUBO solution maps to a somewhat worse Bethel outcome in this instance is consistent with the surrogate being locally informative near its optimum, though this study does not establish that the relationship holds more broadly. The two runs on real quantum hardware fall much further short: IBM's QAOA returns 262 and the QCi Dirac-3 returns 286.

\begin{figure}[htbp]
\centering
\includegraphics[width=0.92\textwidth]{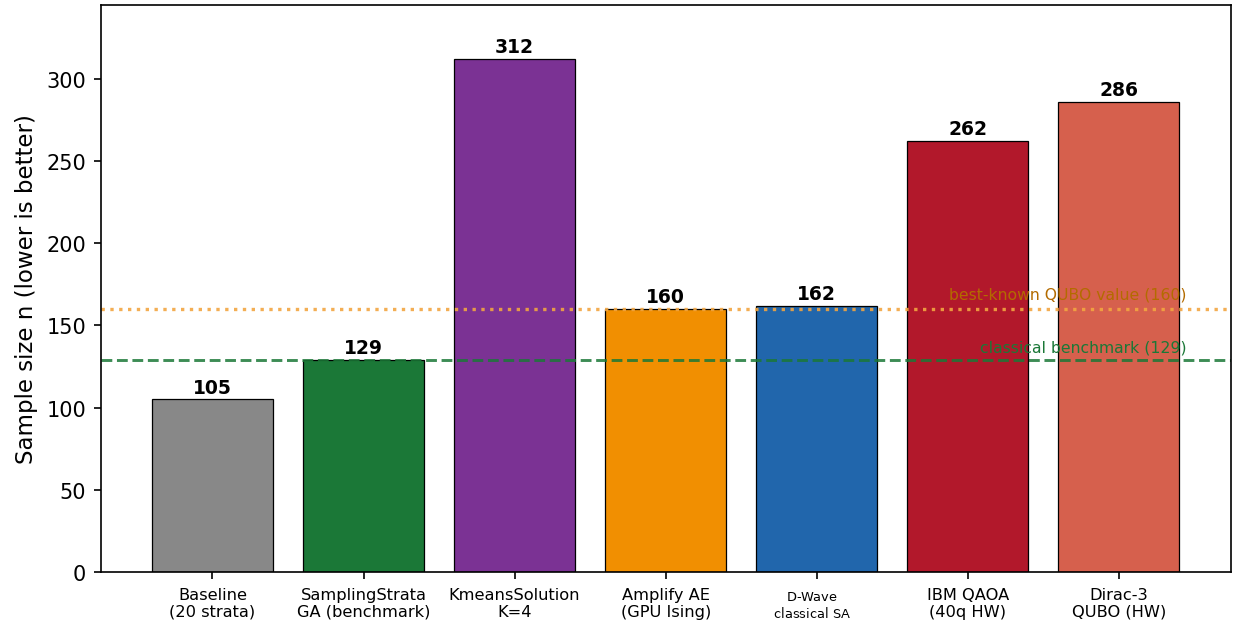}
\caption{Sample size by method on the common twenty-stratum frame (K = 4, CV = 5\%). Lower is better. The dashed green line marks the classical benchmark of 129; the dotted orange line marks the Bethel sample size, 160, associated with the best-known QUBO solution found in this study, whose QUBO objective value is 501.0, reached by Fixstars Amplify AE and confirmed by an independent search.}
\label{fig:results}
\end{figure}

\subsection{Two distinct modes of poor performance in the hardware runs}
\label{sec:two-modes}
It must be stressed at the outset that all the quantum and quantum-inspired routes optimise the same underlying population-weighted dispersion criterion, but not always through the same literal polynomial encoding. D-Wave, Dirac-3 and Amplify AE use the common one-hot QUBO, whereas the IBM implementation uses a binary-label Hamiltonian containing both two-body and four-body terms. None of these routes optimises the Bethel sample size directly, and none has privileged access to the true survey-design objective. The differences observed below therefore reflect solver performance and, in IBM's case, the behaviour of a different encoding of the same underlying surrogate criterion.

It should be stated plainly what the IBM run was and was not. With only two COBYLA iterations, imposed by the free-tier time budget, this is a severely truncated proof-of-execution run on real IBM hardware, not a general measurement of QAOA performance; a properly converged run, as discussed in Section~\ref{sec:ibm-paid}, would need far more iterations. Within that scope, the run failed at the optimisation itself.

The forty-qubit binary-label encoding produced, after transpilation to the heavy-hex lattice, a circuit of depth roughly 4,900 with about 5,300 two-qubit gates, the SWAP overhead of a dense interaction graph being the main contributor. At that depth the measured cost energy sat at zero throughout the short variational loop, and the final bitstring distribution showed no outcome repeated more than once out of the shots taken. As shown exactly in Section~\ref{sec:ibm}, this near-zero energy is precisely what the identity-shifted Hamiltonian implemented here predicts for a broad, near-uniform output, so it requires no separate offset or scaling explanation; the diagnostic evidence that the run failed to optimise lies instead in the absence of any energy improvement across the two iterations and in the diffuse, non-repeating bitstring distribution itself, both consistent with a signal destroyed by accumulated gate error at this circuit depth.

No separate check against an ideal or lightly-noised simulator was performed for this specific run to isolate the two. To give this comparison a quantitative anchor, 10,000 uniformly random four-cluster assignments of the same twenty atomic strata were generated, conditional on all four cluster labels being represented (resampling on the rare occasions, about 1.4\% of draws, that a label was missing, so that the comparison uses the same admissible space, non-empty final strata, as every solver output in this paper), and evaluated with the same Bethel-Chromy allocator: the resulting sample sizes have a median of 314 and range from 162 to 383. IBM's value of 262 lies at approximately the 12th percentile of this distribution of individual random assignments. It is therefore better than a typical single random partition, but remains within the range readily attainable by random search; the number of distinct bitstrings actually sampled during the two COBYLA iterations was not recorded, so a matched best-of-m random comparison could not be constructed, and no such comparison is offered here. The four-cluster partition that was decoded is consistent with a modest best-of-random search rather than with a successful QAOA optimisation, obtained from a run too short and too deep to constitute a meaningful optimisation, rather than as a demonstration that IBM QAOA cannot solve this problem under any conditions.

The Dirac-3 run, by contrast, shows evidence of a non-trivial optimisation process: its final native energy, as reported by the device, was clearly negative, it returned ten distinct low-energy solutions with only a single, easily repaired constraint violation, and its all-to-all photonic connectivity avoids the depth blow-up that crippled the gate-based circuit, since there is no transpilation, no SWAP overhead and no variational loop; the native energy scale is not documented well enough, however, to certify from it alone that the underlying search was effective, as opposed to merely well-behaved. No relationship of the form $E_{\mathrm{native}}=a\cdot E_{\mathrm{common}}+b$ between the device's own reported energy and the common one-hot QUBO energy used elsewhere in this paper is documented by the vendor or established in this study; the two scales are therefore not compared directly anywhere in this paper, and every comparison of Dirac-3 to the other solvers instead uses either the common QUBO energy, computed independently as below, or the downstream Bethel sample size. Given the best-known solution of the QUBO established in this study, from Amplify AE and confirmed by an independent verification search, it is clear that the Dirac-3 did not reach it. Reevaluating the decoded Dirac-3 partition directly under the identical common one-hot objective (same weights, same penalty) used for D-Wave and Amplify AE gives an energy of 936.61, against the best-known 501.0 and D-Wave's 566.0: nearly double the best-known value, confirming quantitatively, not only through the downstream Bethel comparison (286 against the best-known 160), a substantially higher value of the common feasible objective than the value reached by the classical GPU solver in seconds. It searched the correct surrogate landscape but returned a substantially weaker feasible solution.

\subsection{Decomposing the gap: the cost of the formulation and the cost of the solver}
\label{sec:decomposing}
All four surrogate-based routes minimise the same underlying dispersion criterion, although IBM uses a binary-label Hamiltonian rather than the common one-hot QUBO. The distance between each result and the classical benchmark can therefore be discussed in terms of two components: a formulation component, arising from the replacement of the Bethel objective by the dispersion surrogate, and a solver/platform component, arising jointly from hardware, solver dynamics, computational budget, constraint handling and, for IBM, the encoding itself, all of which differ across the four solvers as noted at the outset of this section. This decomposition is empirical rather than certified: both components rest on the best solutions actually found, not on proofs of optimality. The first is the cost of the formulation: the price paid for replacing the Bethel sample size with a clustering surrogate. Estimating it requires the best solution to the surrogate that can be found, independently of any hardware limitation, and this is what Amplify AE and an independent, more thorough simulated-annealing check both supply, converging on an objective value of 501.0 in seconds. This best-known solution corresponds to a sample size of 160, against the classical benchmark of 129, so the formulation is associated with a downstream gap of about twenty-four percent in this instance, reported as the best-known Bethel value associated with the surrogate's best-known solution, not a proven lower bound.

It remains conceivable, though unobserved here, that a different QUBO solution of similar or slightly higher energy could correspond to a lower Bethel sample size, since the surrogate and the true objective are not guaranteed to share the same ordering away from their respective optima.

The second is the solver/platform component: how far a given device, together with everything specific to how it was run, falls short of that best-known value. The two hardware runs can then be assessed. Dirac-3 and Amplify AE use the same one-hot variables and the same dispersion objective, and their decoded feasible partitions are re-evaluated under the same common objective, though their internal constraint handling and search landscapes are not identical. The Dirac-3 returns 286 rather than 160: the platform, solver dynamics and constraint handling differed, while the feasible dispersion objective and final Bethel evaluation were held fixed. IBM's 262 should not be read as a shortfall of this kind, and certainly not as an outcome better than the Dirac-3's: as discussed above, the run performed essentially no optimisation, so its sample size mostly reflects what a near-random four-cluster partition of twenty strata happens to cost. The instructive comparison is that a five-second run on a graphics processor, available at no cost, reaches a substantially better solution of the same problem than either quantum device managed with real hardware time on real machines.

Taken together the two components explain the whole ordering observed here. The surrogate keeps the best solutions found at or above roughly 160, a figure no solver in this study bettered; the classical and quantum-inspired solvers reach it or close to it, the two quantum devices do not. What separates the best-known QUBO result from the classical benchmark is attributed to the formulation, and what separates the quantum devices from that best-known result is attributed to the solver/platform component, on the evidence gathered in this particular experiment; this observed gap is consistent with a substantial formulation component, although residual heuristic error cannot be excluded.

\subsection{Why the surrogate limits the attainable result}
\label{sec:why-surrogate}
As shown in Section~\ref{sec:quantum}, the Bethel--Chromy sample size cannot be represented directly as a compact quadratic function of the assignment variables. The quantum and quantum-inspired solvers therefore minimise an unnormalised population-weighted pairwise dispersion criterion rather than the true survey-design objective.

The results in Section~\ref{sec:decomposing} show the practical consequence of this distinction. Even when the surrogate is minimised effectively by a classical GPU solver, the resulting partition requires a Bethel sample size of about 160, compared with 129 for the genetic algorithm that evaluates the true objective directly. This observed gap is consistent with a substantial formulation component, although residual classical heuristic error cannot be excluded.

The surrogate is not necessarily uninformative: lower dispersion values can still correspond to better stratifications, and the best QUBO solutions found here are substantially better than random or poorly optimised partitions. However, the surrogate and the Bethel objective are not guaranteed to rank candidate partitions in the same order or to share the same minimiser. In particular, the omission of assignment-dependent normalisation terms can favour partitions that are homogeneous according to the pairwise criterion but inefficient under the subsequent sample-allocation problem.

Bringing the optimisation closer to the Bethel objective would require a more strongly hybrid approach. Possible directions include constructing local quadratic approximations from classical Bethel evaluations, using constrained or nonlinear hybrid models, or introducing auxiliary variables to approximate assignment-dependent quantities. Each option increases either the number of variables, the approximation error, or the amount of classical computation surrounding the quantum solver.

The main methodological implication is therefore that improving the hardware alone is not sufficient. Before quantum optimisation can become competitive for optimal stratification, the objective encoded in the solver must reproduce more closely the ranking induced by the Bethel--Chromy allocation.

\section{Beyond the free tier: what paid access would, and would not, change}
\label{sec:beyond-free-tier}
Every limitation encountered in this study was met on the freely available tier of each platform. This section asks how much of the picture is an artefact of that choice, keeping to the structural argument for each platform; commercial details (pricing, quotas, credits) are given in Appendix~\ref{app:paid-access}.

\subsection{D-Wave: paid access buys the QPU, but embedding is likely to constrain scaling}
\label{sec:dwave-paid}
A paid licence would grant access to the real annealer, and it is tempting to assume that the several thousand physical qubits of an Advantage-class machine would accommodate a much larger frame. This is unlikely to hold without qualification. The qubits of an annealer are not fully connected (fifteen neighbours per qubit on Pegasus, twenty on Zephyr), whereas our clustering QUBO, though dense, is not complete: a variable $x_{hk}$ interacts with $(L-1)+(K-1)=22$ others at $L=20$, $K=4$, giving $K\binom{L}{2}+L\binom{K}{2}=880$ edges against $\binom{LK}{2}=3{,}160$ for a complete graph on the same 80 variables, a density of about 28\%.

No embedding of this actual, sparser graph was computed in this study. What is documented is the capacity for embedding a \emph{complete} graph on Pegasus, about 150 variables with perfect yield, degrading in practice beyond roughly 120 owing to broken chains. Applying this complete-graph figure to the one-hot QUBO ($N\times K$ variables) gives an illustrative, conservative reference of roughly thirty to thirty-seven atomic strata at $K=4$; because the true graph is markedly sparser, this is a lower bound on what paid access could plausibly reach, not a prediction of it. D-Wave's hybrid solvers, also paid, accept far larger problems but decompose them classically and call the QPU only on sub-problems, so the genuinely quantum share shrinks accordingly.

\subsection{IBM: paid access buys time, not depth}
\label{sec:ibm-paid}
The IBM free plan's time and session restrictions are why the variational loop had to be cut to two COBYLA iterations. A paid plan would remove exactly these restrictions, providing far more QPU time and error-mitigation techniques, and would allow a fair test of QAOA rather than the truncated one performed here. It would not, however, remove the binding constraint, because that constraint is physical rather than contractual: the transpiled circuit for this problem contains roughly five thousand two-qubit gates, and the error rate per two-qubit gate on current superconducting hardware is such that the accumulated error is expected to degrade the useful signal severely at that depth, independently of how many optimisation iterations are run. Error mitigation extends the usable depth but at exponentially growing sampling cost, so it postpones the limitation rather than removing it. Only substantially lower gate error, genuine error correction, an alternative encoding needing fewer long-range interactions, or a shallower circuit from better compilation would change this conclusion; paid access alone would not.

\subsection{Dirac-3: paid access lifts the binding constraint}
\label{sec:dirac-paid}
The Dirac-3 is the platform on which paid access would help most, since the limit met here (100 decision variables) was an explicit free-tier quota rather than a device property. The device itself is bounded by its total qudit-level capacity, documented as 954 levels in the most specific and current source available \citep{qci2026eqcdirect} (see Appendix~\ref{app:dirac-discrepancy} for a discussion of a second, slightly different figure elsewhere in QCi's own documentation). A binary variable consumes two levels, so full access would accommodate roughly 119 atomic strata at $K=4$, close to the original 164-stratum frame and the single largest gain of the three platforms.

\subsection{What would actually change in the results}
\label{sec:what-would-change}
Combining the three, a common frame of at least roughly thirty atomic strata appears plausible from the D-Wave and Dirac-3 perspectives; IBM's circuit depth, however, would remain a serious limitation and would worsen further at that size, and is not relieved by paid access to any other platform. Such an experiment would sharpen the measurement of the solver/platform component of the gap identified in Section~\ref{sec:decomposing}, but it would not touch the formulation component: the best solution to the clustering QUBO found in this study, by a free classical GPU solver, already lands at about 160 against a classical benchmark of 129, and paid access would not be expected to remove that gap, since every solver would still optimise the same surrogate rather than the Bethel objective. The decisive research problem is therefore not access to bigger machines; it is the construction of a quantum-compatible objective whose optimum coincides more closely with that of the Bethel sample size, along the lines discussed in Section~\ref{sec:why-surrogate}.

\section{Implications and future experiments}
\label{sec:wider-landscape}
The IBM failure documented in Section~\ref{sec:two-modes} appears to have been driven in large part by SWAP routing rather than by the abstract algorithm: a dense QUBO demands interactions between qubits that are not neighbours on the heavy-hex lattice, and the transpiler must shuttle states across the chip, inflating the circuit to roughly 5,300 two-qubit gates. Trapped-ion processors do not have this specific problem: their two-qubit interactions are mediated by the collective vibrational modes of the ion chain, giving effectively all-to-all connectivity with no SWAP overhead \citep{pino2021}, a property independently confirmed in a large-scale, multi-vendor benchmarking study \citep{montanez2025}. This removes the SWAP-routing component of the IBM failure specifically, not necessarily every other contributor to circuit depth, since native gate set, four-body-term decomposition and gate scheduling still affect the final depth on a real trapped-ion device.

This hypothesis was tested offline, for free, before any cloud account was created, by transpiling the identical QAOA circuit against a heavy-hex coupling map (matching IBM) and a fully connected one (matching a trapped-ion topology), using the standard Qiskit transpiler.

\begin{figure}[htbp]
\centering
\includegraphics[width=0.92\textwidth]{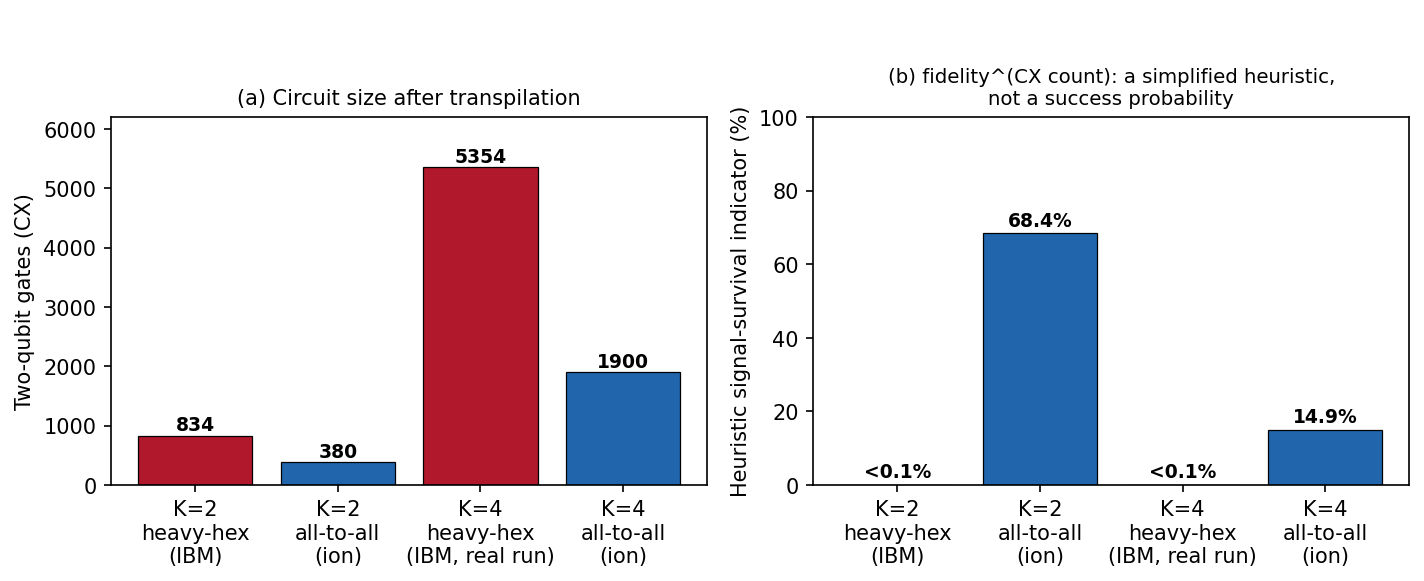}
\caption{Two-qubit gate count and a heuristic signal-survival indicator for the same QAOA circuit, transpiled against a sparse heavy-hex coupling map and against a fully connected one, at K = 2 (20 qubits) and K = 4 (40 qubits). The indicator is $(\text{two-qubit gate fidelity})^{\text{number of CX gates}}$, using 99\% for superconducting and 99.9\% for trapped-ion hardware; it is a simplified proxy, not a computed success probability.}
\label{fig:swap}
\end{figure}

At K = 4, routing inflates the gate count from 1,900 (all-to-all) to 5,354 (heavy-hex), the latter matching almost exactly the 5,354 two-qubit gates actually measured on the real IBM run, a useful check that the offline measurement matches reality; at K = 2 the same effect more than doubles the gate count (380 to 834). Translating gate count into a heuristic signal-survival indicator, $S_{\mathrm{heuristic}}=f_2^{\,G}$ (a representative two-qubit gate fidelity raised to the gate count, not a rigorous success probability), the heavy-hex configuration falls to well under 0.1\% at both K values, consistent with the flat, energy-zero outcome observed on IBM, while the all-to-all configuration retains a non-negligible value (about 15\% at K = 4, and about two-thirds at K = 2). This supports, without by itself proving, the connectivity-based explanation of the IBM failure, and it identifies K = 2 on a trapped-ion device as the configuration where connectivity should matter most: a concrete, falsifiable, low-cost prediction.

The recommended next experiment is therefore to recompile the same logical Hamiltonian and ansatz (the parameterised circuit structure optimised variationally) for a fully connected coupling map on a trapped-ion device, IonQ Forte or Quantinuum H2/Helios being the relevant machines at this scale \citep{ionq2026azure,quantinuum2026systems}, starting with the cheaper K = 2 case as a minimal-cost first test of the connectivity hypothesis before attempting the full K = 4 problem. This would not, however, change the conclusion regarding the formulation established in Section~\ref{sec:decomposing}: no solver in this study, on any platform, reached a Bethel value below about 160 while minimising this QUBO, so narrowing that gap remains the more consequential open problem.

Several other platforms were considered and set aside, for reasons ranging from closed access (Google Willow) to structural mismatch with the present dense clustering formulation (neutral-atom devices) to redundancy with the measurement already obtained from Amplify AE (other dense-QUBO machines); a full account, together with a summary table of every platform considered, is given in Appendix~\ref{app:platforms}.

\section{Conclusions}
\label{sec:conclusions}

On the optimal-stratification problem studied here, at the scale accessible on freely available hardware, classical and quantum-inspired methods clearly outperform the gate-based and photonic devices tested. The classical genetic algorithm sets the benchmark at a sample size of 129. The best-known QUBO solution found in this study has an objective value of 501.0, established by a free GPU Ising machine and independently confirmed by one further classical search; evaluated with the classical Bethel-Chromy allocator, it maps to a sample size of 160. A third, weaker classical search on the same QUBO landed on a visibly worse local optimum, mapping to 162. IBM's gate-based processor and the vendor-described photonic entropy-quantum-optimisation device (Dirac-3) do not reach even that, trailing at 262 (IBM) and 286 (Dirac-3).

The single most useful addition to the experiment was the solver that is not quantum at all. Fixstars Amplify AE removes the principal quantum-specific hardware limitations considered here at once: it accepts the dense QUBO natively, so there is no minor-embedding and no chain of physical qubits; it needs no transpilation, so there is no SWAP overhead; and it is classical, so there is no gate noise. What remains, once all of that is removed, is an empirical combination of formulation mismatch and residual classical heuristic error, without the quantum-specific hardware effects present on the physical devices; no solver in this study, quantum or classical, found a Bethel value below roughly 160 while minimising the surrogate. That figure is the best Bethel value associated with a QUBO solution found in this study, and it lies twenty-four percent above the classical benchmark. This makes the diagnosis reasonably clear, and it separates two obstacles that are easily conflated, while stopping short of a formal proof for either. The first is the surrogate. The QUBO minimises within-cluster dispersion, not the Bethel sample size, and the two are not guaranteed to share an optimum; no solver tried in this study closed that gap while minimising this objective, though this is empirical evidence rather than a mathematical guarantee that none could. The second is the solver and platform component. Both quantum devices fell well short of even the best-known surrogate optimum, for reasons that are now reasonably well characterised: the gate-based IBM circuit, inflated by SWAP routing on a sparse heavy-hex lattice to roughly five thousand two-qubit gates and run for only two truncated optimisation steps, returned a result consistent with a modest best-of-random outcome rather than with successful QAOA optimisation; the photonic Dirac-3, free of that specific problem thanks to all-to-all connectivity, did optimise but settled far above the value that a free graphics-processor solver reaches in seconds.

Two structural obstacles would therefore have to be addressed before quantum optimisation could become competitive for optimal stratification, and they should be addressed in that order. First, the objective actually solved should be brought closer to the Bethel sample-size objective and away from a simple homogeneity proxy, along one of the lines discussed in Section~\ref{sec:why-surrogate}; without this, the best result found in this study for any solver minimising the QUBO remains 160 against a classical benchmark of 129. Only once that gap is narrowed does hardware quality become the binding question. Until both are addressed, he practical recommendation is therefore to continue using the SamplingStrata genetic algorithm as the method of choice for this problem.

Section~\ref{sec:wider-landscape} identifies the one experiment still worth running. Because the gate-based failure was very plausibly aggravated by SWAP routing on a sparse lattice, the same circuit should be executed on a trapped-ion machine, whose all-to-all connectivity removes that overhead entirely; this would help separate a general limitation of gate-based quantum computing from a specific limitation of sparse hardware.

It would not, however, change the conclusion above regarding the formulation, since no solver in this study minimising this particular QUBO reached a Bethel value below about 160. The value of the exercise is therefore methodological rather than operational. It shows how the optimal-stratification problem can be cast in QUBO form, it identifies which stages of each workflow are quantum and which remain classical, and, by including a purely classical solver as a control, it separates, as far as the evidence gathered here allows, the cost of the formulation from the cost of the solver and platform, which is a useful starting point for any assessment of quantum advantage on a real survey-design task.

In conclusion, the comparison shown in this paper is best read as a study of objective mismatch plus computational error. SamplingStrata searches a difficult landscape but evaluates the correct target at every step. The QUBO routes simplify the landscape into a hardware-compatible proxy, then incur an additional solver-specific approximation. The experiment shows that both layers matter, with the proxy mismatch already creating a substantial gap.

\appendix

\section{Appendix A - Extended mathematical derivations}
\label{app:math}
This appendix gives, in full, several derivations that the main text states as results only, for length. Notation follows the main text throughout.

\subsection{Full expansion of the one-hot penalty (Section~\ref{sec:constraints})}
\label{app:math-penalty}
Because $x_{hk}^2=x_{hk}$ for binary variables, the one-hot penalty can be expanded as
$${\left(1-\sum_{k=1}^{K} {x}_{hk}\right)}^{2}=1-\sum_{k=1}^{K} {x}_{hk}+2\sum_{1\le k<l\le K} {x}_{hk}{x}_{hl}.$$
Substituting into $Q_{\mathrm{pen}}(X)=\lambda\sum_h(1-\sum_k x_{hk})^2$ and adding the dispersion term $Q_{\mathrm{disp}}(X)$ gives, apart from the constant $\lambda L$, the complete quadratic polynomial quoted in the main text,
$$Q\left(X\right)=-\lambda\sum_{h=1}^{L} \sum_{k=1}^{K} {x}_{hk}+\sum_{k=1}^{K} \sum_{1\le h<h'\le L} {w}_{hh'}{x}_{hk}{x}_{h'k}+2\lambda\sum_{h=1}^{L} \sum_{1\le k<l\le K} {x}_{hk}{x}_{hl}+\lambda L.$$
After stacking the $LK$ binary indicators into a vector $x$, this is the standard QUBO form $\min_{x\in\{0,1\}^{LK}} x^TQx$, where the diagonal elements of $Q$ contain the linear coefficients and the off-diagonal elements contain both the within-stratum dissimilarities and the one-hot penalties.

\subsection{QUBO-Ising equivalence (Section~\ref{sec:notions})}
\label{app:math-ising}
The Ising representation expresses the QUBO energy through spin variables $s_i\in\{-1,+1\}$ rather than binary variables $x_i\in\{0,1\}$, with linear terms describing single-spin contributions (main effects, in statistical language) and quadratic terms describing pairwise interactions. The two representations are equivalent under $s_i=1-2x_i$, equivalently $x_i=(1-s_i)/2$: substituting this relation into the QUBO changes the coefficients and may add an irrelevant constant, but preserves the ordering of configurations and the set of minimisers, since the transformation is a strictly monotonic (in fact affine) reparameterisation applied identically to every configuration.

\subsection{Pairwise identity connecting the surrogate to the Bethel objective (Section~\ref{sec:relationship})}
\label{app:math-pairwise}
The identity
$$\sum_{h:{z}_{h}=k} {N}_{h}{\left({\overline{Y}}_{hj}-{\overline{Y}}_{kj}\right)}^{2}=\frac{1}{2{N}_{k}\left(z\right)}\sum_{h=1}^{L} \sum_{h'=1}^{L} {N}_{h}{N}_{h'}{\left({\overline{Y}}_{hj}-{\overline{Y}}_{h'j}\right)}^{2}{x}_{hk}{x}_{h'k}$$
follows from the standard decomposition of a weighted sum of squared deviations from the mean into a sum over all pairs: expanding the right-hand side's squared difference and using $\sum_h N_h x_{hk}\overline{Y}_{hj}=N_k(z)\overline{Y}_{kj}(z)$ and $\sum_h N_h x_{hk}=N_k(z)$ recovers the left-hand side after cancellation. For a fixed value of $N_k$, the numerator on the right-hand side is quadratic in the assignment indicators; because $N_k(z)$ itself varies with the candidate partition, dropping the factor $1/\{2N_k(z)\}$ to obtain a valid QUBO changes the criterion and can alter the ranking of partitions, especially when final-stratum sizes differ substantially.

\subsection{Exact expectation of the implemented IBM Hamiltonian (Section~\ref{sec:ibm})}
\label{app:math-ibm-energy}
Dropping the identity term from the cost Hamiltonian $H_C=\sum_{h<h'}w_{hh'}\delta(z_h,z_{h'})$ gives the operator actually implemented,
$$H_C^{\mathrm{impl}}=H_C-\tfrac{1}{4}\sum_{h<h'}w_{hh'}I,$$
a constant shift lowering every eigenvalue uniformly. For uniformly random labels with $K=4$, $P(z_h=z_{h'})=1/4$ for any pair $h\neq h'$, so $E[\delta(z_h,z_{h'})]=1/4$ under the untranslated Hamiltonian; after the identity term is removed, the expected cost of a uniform distribution over the four labels is therefore exactly zero,
$$E\left[H_C^{\mathrm{impl}}\right]=\sum_{h<h'}w_{hh'}\left(\tfrac{1}{4}-\tfrac{1}{4}\right)=0,$$
by construction, not as an empirical coincidence requiring a separate explanation such as an offset or scaling error. The near-zero measured energy on the real IBM run is therefore exactly what the implemented Hamiltonian predicts for a broad, near-uniform output distribution; the evidence that the run failed to optimise lies in the absence of any energy improvement over the two iterations and in the diffuse, non-repeating bitstring distribution actually observed, not in the zero energy value taken on its own.

\subsection{QAOA circuit structure and run parameters (Section~\ref{sec:ibm})}
\label{app:math-qaoa-params}
QAOA searches the cost landscape through a parameterised quantum circuit. The circuit starts from a superposition of possible binary labels and alternates two operations: evolution under the cost Hamiltonian, which assigns lower energy to assignments with lower surrogate cost, and evolution under a mixing Hamiltonian, which allows the circuit to explore alternative assignments. For a circuit with $p$ layers, these operations are repeated $p$ times, their duration controlled by a set of variational parameters; the quantum processor prepares and measures the corresponding state, while a classical optimiser (COBYLA in this study) updates the parameters to reduce the estimated expected energy.

The real hardware run used the ibm\_kingston backend (Heron r2, 156 qubits), shown in Figure~\ref{fig:ibm} of the main text. The exact QAOA layer count $p$ and the number of shots per circuit evaluation were not recorded in a form that can be reported precisely here; the final partition was the lowest-cost bitstring observed among those sampled during the two COBYLA iterations. This gap affects only the precise reproducibility of this one run, not the conclusions drawn from it, which rest on the measured circuit depth and gate count rather than on these parameter values.

\subsection{D-Wave annealing schedule (Section~\ref{sec:dwave})}
\label{app:math-dwave-schedule}
The search begins from the ground state of a simple transverse-field Hamiltonian $H_0$ and evolves according to the time-dependent Hamiltonian
$$H\left(t\right)=A\left(t\right){H}_{0}+B\left(t\right){H}_{P}, \quad 0\le t\le T,$$
where $t$ denotes the progression of the annealing process. At the beginning, the transverse-field term dominates and allows the system to explore alternative configurations; toward the end, the problem Hamiltonian $H_P$ dominates, so lower-energy configurations correspond to better values of the QUBO objective. A further implementation step is required because the logical Ising model is dense, whereas the QPU has a fixed sparse connectivity graph: minor embedding represents a logical variable by a chain of connected physical qubits, so the number of physical qubits can be substantially larger than the number of logical variables, and this can affect the quality of the search through chain length, chain strength, broken chains, coefficient scaling and hardware noise.

\FloatBarrier

\section{Appendix B - Extended discussion: what paid access would change}
\label{app:paid-access}
This appendix expands Section~\ref{sec:beyond-free-tier} of the main text with the commercial and quota detail omitted there.

\subsection{D-Wave}
On the free plan the QPU solvers were not authorised at all, which is why the D-Wave QUBO was solved in this study with classical simulated annealing. D-Wave's hybrid solvers, available on paid plans, accept problems with up to a million variables and would swallow the original 164-stratum frame without difficulty; but they operate by classically decomposing the problem and invoking the QPU only on sub-problems, so the genuinely quantum share of the computation shrinks accordingly, and one is largely measuring a classical solver again.

\subsection{IBM}
The IBM free plan grants ten minutes of QPU time in a rolling twenty-eight-day window and forbids session mode, which is why the variational loop in this study had to be cut to two COBYLA iterations, each queued as an independent job. A paid plan removes exactly these restrictions: it provides far more QPU time, dedicated sessions that eliminate the queue between iterations, and access to error-mitigation and error-suppression techniques, so that the variational loop could run to convergence with dozens or hundreds of iterations and far more shots per circuit. As discussed in the main text, this would not remove the circuit-depth limitation, which is physical rather than contractual.

\subsection{Dirac-3: the 949 versus 954 discrepancy}
\label{app:dirac-discrepancy}
The free tier rejects any degree-two problem with more than one hundred decision variables, which at $K=4$ caps the frame at twenty-five atomic strata. The physical device is bounded instead by its total qudit-level capacity. QCi's own documentation is not perfectly consistent on this figure: the Dirac-3 Developer Beginner Guide states 949, while the eqc-direct integer-solver documentation states 954 and gives, as a worked example, a maximum of 477 binary variables, consistent with 954 rather than 949 (which would give 474). Neither source was the specific endpoint queried for the free-tier rejection reported in the main text, which returned only the separate 100-variable limit. Lacking a documented figure for the exact endpoint relevant to this study, the main text adopts the eqc-direct figure of 954 as the more specific and more recently verified of the two, giving roughly 119 atomic strata at $K=4$ under full paid access.

\subsection{Trapped ions: access and cost}
Access to trapped-ion hardware (IonQ, Quantinuum) is through Amazon Braket or Microsoft Azure Quantum, both pay-per-use. Microsoft has historically advertised an automatic credit of USD 500 per participating hardware provider, and up to USD 10,000 through its Research Credits programme; these specific figures do not appear on the current Azure Quantum pricing page, and their continued availability, amount and eligibility depend on the specific provider, region, subscription plan and time period, change without notice, and should be re-verified directly with Microsoft before being relied upon. A word of caution on cost: IonQ bills per gate-shot, so cost depends on the number of shots and on whether error mitigation is enabled, neither of which is fixed here; no explicit cost estimate is offered for the $K=4$ circuit, with its roughly two thousand two-qubit gates, beyond noting that it is the more expensive of the two options by a wide margin. The $K=2$ encoding, at twenty qubits and 380 two-qubit gates on a fully connected layout, is the cheaper test and the one recommended in the main text as the first, minimal-cost step, though its exact cost was likewise not computed here.

\FloatBarrier

\section{Appendix C - Platforms considered but not used}
\label{app:platforms}
The four solvers used in the main study were chosen because they were freely accessible, not because they are the only options. This appendix surveys the platforms that were considered but not used, and explains why.

\subsection{Google: effectively closed}
The obvious omission is Google. As of access in July 2026, its hardware is not generally available: the Quantum Computing Service has remained a limited preview granting time to selected research partners rather than open self-service, so one cannot simply register and run circuits on the Willow processor, in contrast with IBM's open cloud \citep{google2026}. Google had opened a Willow Early Access Program inviting research proposals; that route was by competitive selection, with a proposal deadline of 15 May 2026, and the programme page now indicates that selected applicants have been notified, at least one such selection (a King's College London team) having been publicly announced in late May 2026. The programme was, in any case, aimed at experiments designed specifically to probe the device rather than at applied optimisation studies, and no application was made for the present study. This is a time-sensitive, commercial detail that should be re-verified against Google's current documentation before being relied upon; for the purposes of a study like this one, at the time of access, Google was not an option.

\subsection{Other dense-QUBO machines, and why Amplify AE was enough}
The family of machines purpose-built for densely connected QUBOs is broader than the single member used in the main study. The Fujitsu Digital Annealer solves fully connected QUBO problems on application-specific CMOS hardware, with an annealing algorithm using parallel-trial and dynamic-escape mechanisms \citep{aramon2019}, and its fully connected architecture is aimed precisely at problems that are hard for both conventional computers and quantum annealers. Toshiba's SQBM+, built on the Simulated Bifurcation algorithm \citep{goto2019}, and NEC's Vector Annealing service occupy the same niche.

Neither was run, for a practical and a scientific reason. Practically, neither is free: SQBM+ requires a separate subscription with its own endpoint, and the Digital Annealer a Fujitsu licence, whereas Amplify AE is available at no cost. Scientifically, they would add comparatively little on their own, because Amplify AE has already provided the key measurement: the best-known QUBO solution found when quantum-specific hardware limitations are absent and a strong classical heuristic is used, corroborated by one further independent simulated-annealing search. A second or third dense-QUBO machine could provide an additional independent comparison with this best-known result, but would be unlikely to change the qualitative picture already established.

One product deserves a word only to set it aside. Fixstars also markets Amplify SE, the Scheduling Engine, which is sometimes mistaken for a variant of AE. It is a different tool entirely: a domain-specific solver for scheduling problems, in which the user declares jobs, machines and tasks with their processing times, and the engine minimises the makespan. It deliberately requires no mathematical formulation from the user and does not accept a QUBO at all. Optimal stratification is not a scheduling problem, so Amplify SE has no bearing on it.

\subsection{What would not help}
Two other families were not pursued, for reasons specific to the present formulation rather than general unsuitability. Neutral-atom machines, such as QuEra's Aquila and Pasqal's Fresnel, are analog devices whose natural problem is the maximum independent set on unit-disk graphs; they were not pursued because the present dense clustering formulation does not map naturally to the native analog problem classes available on those platforms without substantial reformulation. The remaining gate-based superconducting platforms, Rigetti and IQM, share the same sparse fixed-layout connectivity that contributed to the IBM failure; they would not isolate the role of connectivity as cleanly as an all-to-all trapped-ion experiment, because they also rely on sparse fixed-connectivity layouts, though they could still differ from IBM in fidelity, native gate set, compilation and calibration, so a similar outcome is expected rather than certain.

\subsection{Summary table}
\begin{table}[htbp]
\centering
\small
\begin{tabular}{@{}p{2.6cm}p{2.4cm}p{2.6cm}p{4.6cm}@{}}
\toprule
\textbf{Platform} & \textbf{Type} & \textbf{Status} & \textbf{Reason} \\
\midrule
Fixstars Amplify AE & GPU Ising engine & USED (n = 160) & Free Basic plan; 8,192-bit fully-connected cap, our 80 variables well within it \\
D-Wave QPU & Quantum annealing & NOT USED & Not authorised on the free plan \\
D-Wave (Ocean classical SA) & Classical SA, same pipeline & USED (n = 162) & Ran in place of the QPU, on the free plan \\
IBM Quantum & Gate-based, sparse & USED (n = 262) & Severely truncated run at depth \textasciitilde{}4,900 \\
QCi Dirac-3 & Photonic, all-to-all & USED (n = 286) & Did not reach the best-known QUBO value found in this study \\
Quantinuum / IonQ & Trapped ion, all-to-all & NOT USED - worth running & No SWAP overhead; paid, credits available \\
Fujitsu DA / Toshiba SQBM+ & Dense-QUBO hardware & NOT USED - redundant & Could provide an additional independent comparison with the best-known result \\
Google Willow & Superconducting & NOT AVAILABLE & Closed; access by proposal only \\
Rigetti / IQM & Superconducting, sparse & NOT USED - less diagnostic for connectivity & Same sparse fixed-layout as IBM; would not isolate connectivity as cleanly as an all-to-all device \\
QuEra / Pasqal & Neutral atom, analog & NOT USED - requires substantial reformulation & Native analog problem class does not match the present dense clustering formulation \\
Fixstars Amplify SE & Scheduling engine & NOT APPLICABLE & Not a QUBO solver at all \\
\bottomrule
\end{tabular}
\caption{Summary of platforms considered in this study.}
\label{tab:landscape}
\end{table}

\bibliographystyle{apalike}
\bibliography{refs}

\end{document}